\documentclass[article,12pt]{article}
\usepackage{latexsym}
\usepackage{epsfig,subfigure, subfloat, graphicx, float }
\usepackage{amsmath} 
\usepackage[colorlinks]{hyperref} 
\usepackage{authblk}
\usepackage[toc,page]{appendix}
\usepackage[colorlinks]{hyperref}
\usepackage[usenames,dvipsnames]{color}
\hypersetup{
     breaklinks=true,
    pdfstartview={FitH},    
    colorlinks=true,       
    linkcolor=blue,          
    citecolor=red,        
    filecolor=magenta,      
    urlcolor=blue,           
    anchorcolor=green,      
    linktocpage=true
}

\def\doi{http://doi.org}

\title{\bf Superradiant instability  and asymptotically AdS hairy black holes in $F(R)$-charged scalar field theory}  \author[1]{A. Rahmani
\thanks{a\_rahmani@sbu.ac.ir}}
\author[1]{M. Honardoost\thanks{m\_honardoost@sbu.ac.ir}}
\author[1]{H. R. Sepangi \thanks{hr-sepangi@sbu.ac.ir}}
\affil[1]{\small Department of Physics, Shahid Beheshti University, G.C., Evin, Tehran
19839, Iran}

\begin{document}
\newcommand\be{\begin{equation}}
\newcommand\ee{\end{equation}}
\newcommand\bea{\begin{eqnarray}}
\newcommand\eea{\end{eqnarray}}
\newcommand\bseq{\begin{subequations}} 
\newcommand\eseq{\end{subequations}}
\newcommand\bcas{\begin{cases}}
\newcommand\ecas{\end{cases}}

\maketitle

\begin{abstract}
We study the phenomena of superradiance for $F(R)$-Maxwell black holes in an AdS space-time. The AdS boundary plays the role of a mirror and provides a natural confining system that makes the superradiant waves bouncing back and forth between the region near the horizon and the reflective boundary, causing a possible superradiant instability. We obtain numerical solutions for static hairy black holes in this scenario and investigate  their instability and explicitly address the stability of such solutions for spherical perturbations under specific conditions for the scalar charge and AdS radius. It is shown that for a small scalar charge or AdS radius the static hairy solution is stable under spherical perturbations. We conclude that under such conditions, new hairy black holes emerge as a possible endpoint of superradiant instability of the system.
\end{abstract}

\section{Introduction}

Black holes as exact solutions of Einstein field equations of General Relativity (GR) have always been regarded as mysterious and fascinating objects and are presently considered  as a prime source of investigation in GR and related subjects \cite{jac,herd}. Just one year after Einstein published his theory of general relativity, Karl Schwarzschild presented the first exact static, spherically symmetry vacuum solution in 1916 \cite{sch} and ever since the task of discovering new exact vacuum
solutions to the field equations has been a challenging and at times a fascinating one with, more often than not, unforeseen consequences \cite{bhsolu}. In particular, the subject of hairy black holes and their instability has been attracting a considerable amount of attention in the recent past.

According to no-hair conjecture introduced by Wheeler and Ruffini, all 4-dimensional stationary, asymptotically flat solutions in Einstein-Maxwell theory are characterized by only 3 parameters, namely mass $M$, angular momentum $J$ and electric charge $Q$ \cite{ruff}. This conjecture means that in GR, black holes are very simple objects and one cannot differentiate between black holes with the same ``Gaussian'' charge. In 1967, a uniqueness theorem of black hole solutions was proposed by Bekenstein and Israel which mathematically supports the no-hair theorem \cite{isra}, that is, the solutions are indistinguishable from each other unless some extra parameters which do not relate to any Gaussian charge are present. Consequently, dropping any initial assumption of the no-hair conjecture may result in hairy black holes \cite{herd} which play an important role in black hole physics \cite{herd,vol} today.

The first example which confirmed the violation of the no-hair conjecture was found by Volkov and Gal'tsov in 1989 \cite{vol,vol2}. They showed that in Einstein-Yang-Mills theory with gauge group $SU(2)$, the additional information, called the primary hair, is the number of oscillations of the Yang Mills field \cite{vol,win,eli}.
It may seem that the presence of additional fields such as a free scalar field, a massive vector  or spinor field may lead to hairy black holes, but a large number of no-hair theorems, mostly by Bekenstein, insist on the absence of any such solutions \cite{vol}, also see \cite{vol,herd,bek} for a review of hairy solutions.

A pioneer in supperradiance investigation was Klein  \cite{pani} when in 1929 found an interesting result of an electron scattering off a potential barrier in the context of Dirac equation \cite{kli}. In non relativistic quantum mechanics, exponential damping is expected due to an electron tunneling through a barrier. However, Klein showed that if the  potential is strong enough, no exponential damping will occur through the transition from the reflecting potential barrier \cite {klei}. This is called the \emph{old-Klein Paradox} which is interpreted today via spontaneous pair production in the presence of strong electromagnetic fields.
However, what we now call superradiance is more close to Zel'dovich's solution which showed that under certain conditions the scattered amplitude from an attractive rotating surface would be amplified \cite{pani}. Therefore by supperradiance, we refer to amplification of scattered (bosonic) test fields from black holes as a dissipative system \cite{pani,brito}. In GR this wave packet amplification may occur if the background is a Kerr or Reissner-Nordstr\"{o}m (RN) black hole \cite{pani,win}. Just as in the Klein case, this amplification is explained trough spontaneous pair production for bosonic fields.
If a superradiant field becomes trapped in an enclosed system, then as a result of energy extraction,  the stability of the background black hole will be uncertain. In this situation the scalar field interacts with the background black hole repeatedly and the scattered wave amplitude increases exponentially which leads to superradiant instability. This type of instability was first investigated by Press and Teukolsky in 1972 \cite{press}. They called this phenomenon the \emph{black hole bomb} which describes a scenario in the context of a flat space-time black hole surrounded by mirrors where a superradiant field is bounced back and forth between the region near
the horizon and a reflective mirror, leading to an eventual exponential growth and instability of the system. They assumed a perfect reflecting mirror to reflect the superradiant field onto the background black hole. The reflecting process is carried out repeatedly until the stability of the system ceases to exist \cite{cardos}.

As was mentioned above, a superradiant field can be trapped in an enclosed system. The question now arises as to what an enclosed system is. To answer this question a reflective wall which can be a fictional mirror \cite{win} or embraces other physical properties of the theory \cite{kerrhod2,kerrhod} was proposed. For example a mass term in the theory of a massive scalar test field near a Kerr black hole can play the role of a reflective mirror \cite{kerrhod2,kerrhod},  whereas such construction is not sufficient for a charged black hole. In other words, the Reissner-Nordstr\"{o}m space-time does not experience superradiant instability due to a massive scalar test field unless one imposes an artificial mirror to enclose the system \cite{masswin,hod,massherd,mass2herd}. In contrast, special characteristic features of some systems like the intrinsic boundary in an AdS space-time can be considered as a natural reflecting mirror for the system \cite{ads1,ads2,ads3,ads4}. Furthermore, in most cases the final fate of superradiant instability is still an open question. A new stable hairy black hole or an explosive phenomenon of bosenova (bose supernova) may be the main plausible endpoint of superradiant instability which have been proposed so far \cite{win,gual,bosch}. These models are extensively explored and often considered in an AdS/CFT context \cite{oscar,oscar1} where we can find a phase transition between the familiar RN-AdS black holes and hairy black holes. As an example, in \cite{win}, the endpoint of  superradiant instability was investigated for a massless scalar field around a charged black hole enclosed by a mirror. It was shown that a static hairy solution may be envisaged as the final state of superradiant instability. To achieve this, spherically symmetric perturbations of hairy solutions have been considered with the result that the hairy black hole is stable at the point of the first zero of the scalar field equilibrium  where the mirror is located \cite{win}.

Standard cosmology has not been quite successful in explaining some of the observational features of the universe, notable amongst which are the late time acceleration,  nature of the cosmological constant (dark energy) and dark matter \cite{fara,felice}.  Such shortcomings has led to a certain class of modifications to GR, what is now referred to as  $F(R)$ gravity where a function of the Ricci scalar $R$ replaces $R$ in Einstein-Hilbert action. In addition, some challenging theoretical  problems relevant to quantum gravity have encouraged the proliferation of $F(R)$ gravity \cite{z} in the recent past.  Also, it is believed that there is a precise formalism for the Newtonian and post-Newtonian limit of $F(R)$ theories in the presence of matter \cite{stable}. For $F(R)$ theories to make sense we need $\frac{dF}{dR}>0$ and $\frac{d^{2}F}{dR^2}>0$, otherwise ghost and tachyonic instability will occur \cite{hendi}. MG has been attracting enormous interest in the recent past which is still going on to this date. Therefore, the study of superradiance phenomena is well warranted within the context of MG.

Following \cite{win}, we study superradiant instability of Maxwell theory in a class of $F(R)$  scenarios and discuss the possible outcome of the instabilities mentioned above. Our numerical solutions are based on the shooting method which generally is a method to find equilibrium static black hole solutions of the field equations, in which the initial conditions are replaced by the boundary conditions at the event horizon. This is in contrast to what is done in \cite{bosch} where numerical time evolutions of the full PDEs are carried out for given initial and boundary conditions at the horizon and AdS boundary. To replace GR with any modified theory, one must first examine the candidate theory from different relevant angles and to this end superradiant instabilities  are no exception. Spherically symmetric solutions of $F(R)$ gravity have been wieldy studied with constant (positive/negative) \cite{Elizal,bhinfr1} and non-constant curvature \cite{sebas}. The modified field equations show that the maximally symmetric solutions of $F(R)$ necessarily correspond to $R=C$, namely de-Sitter ($C>0$) or anti-de Sitter ($C<0$) spaces, just like GR with a cosmological constant $ \Lambda$ \cite{bhinfr, soti}. The presence of AdS solutions in $F(R)$ theory provides a favorable context to study AdS/CFT correspondence \cite{ baz, hart}. So, apart from cosmological necessity mentioned before, our main motivation for investigating the stability of $F(R)$-Maxwell theory is based on AdS/CFT duality. In \cite{moon} exact static spherically symmetric vacuum solutions of $F(R)$-Maxwell theory was analytically derived and later generalized to stationary solutions  where the rotational degrees of freedom were also considered \cite{frkerr}. Furthermore, the stability of Kerr black holes in $F(R)$ theory was investigated in \cite{myu,myun} which showed that superradiant instability will lead to the Kerr black hole instability in $F(R)$ theory.

In this work we introduce two models in conjunction with superradiance phenomena in $F(R)$ gravity. The first model represented by $F(R)=R-\frac{\mu^{4}}{R}$ \cite{felice,hendi,alej,noji} is considered as a toy but useful model. Although, this simple model is free from Dolgov-Kawasaki instability in an AdS space-time (since $\frac{d^{2}F}{dR^{2}}>0$ for $R<0$), it is ruled out because of some cosmological incompatibilities \cite{astro}. Nevertheless, considerations of black hole solutions in such a  model with constant negative curvature is worthwhile and well suited to study the AdS/CFT duality and its consequences. We then move on to study a more realistic model described by $F(R)=R-\lambda e^{-\alpha R}$ which is a reasonable model for cosmology and is compatible with solar system experiments \cite{hendi}. In addition, this model too does not suffer from the Dolgov-Kawasaki instability in AdS space-time. We investigate the static, spherically symmetric solutions of the resulting field equations and proceed to investigate the stability of hairy solutions and superradiant instability.

\section{Preliminaries}

We consider a massless charged scalar field $\Phi$, projected towards a charged, spherically symmetric black hole in a $F(R)$ gravity framework described by the action

\begin{equation}\label{eq1}
  S=\int d^{4}x \sqrt{-g} \left[F(R)-\frac{1}{4}F_{\mu \nu}F^{\mu \nu} -\frac{1}{2}g^{\mu\nu}(D_{(\mu}^* \Phi^* D_{\nu)} \Phi)\right],
\end{equation}
where we set $8\pi G=1$ and $F(R)=R +f(R)$ with a star representing the usual complex conjugation. Also, $F_{\mu \nu}=\nabla_{\mu}A_{\nu}-\nabla_{\nu}A_{\mu} $  and

\begin{equation}\label{2}
  D_{\mu}=\nabla_{\mu}-iqA_{\mu},
\end{equation}
where $q$ indicates the charge of the scalar field.
Variation of the action with respect to dynamical fields leads to equations of motion
\begin{equation}\label{eq3}
 F_{R}R_{\mu \nu}-\frac{1}{2}F(R)g_{\mu \nu}-\nabla_{\mu}\nabla_{\nu}F_{R}+g_{\mu \nu}\Box F_{R} = kT_{\mu \nu} ,
\end{equation}
\begin{eqnarray}
   && \nabla_{\mu}F^{\mu \nu}=J^{\nu}, \label{eq4}\\
   &&D_{\mu}D^{\mu}\Phi=0,\label{eq55}\\
   &&T_{\mu \nu}=T_{\mu \nu}^{\Phi}+T_{\mu \nu}^{F}\label{6},
\end{eqnarray}
where $F_{R}$ represents $\frac{d F(R)}{d R}$ and the field current $J^{\mu}$ and scalar and gauge field parts of the total energy momentum tensor, namely $T_{\mu \nu}^{\phi}$ and $T_{\mu \nu}^{F}$ are defined as
\begin{eqnarray}
   && J^{\mu}=\frac{iq}{2}[\Phi^{*}D^{\mu}\Phi-\Phi(D^{\mu}\Phi)^{*}], \\
   &&T_{\mu \nu}^{\phi}=D^{*}_{(\mu}\Phi^{*}D_{\nu)}\Phi-\frac{1}{2}g_{\mu \nu}[g^{\lambda \eta}D^{*}_{(\lambda}\Phi^{*} D_{\eta)}\Phi],\\
   &&T_{\mu \nu}^{F}=F_{\mu \lambda}F_{\nu}^{\lambda}-\frac{1}{4}g_{\mu \nu}F_{\lambda \eta}F^{\lambda \eta}\label{9}.
\end{eqnarray}

At the linear level, because the background scalar field vanishes, the scalar field can be turned on without back reacting on the electromagnetic fields. In this work we focus on a static, spherically symmetric background with negative constant curvature $R_{0}$ in 4 dimensions so that we deal with an asymptotically AdS space-time according to

\begin{equation}\label{eq10}
ds^{2} =-N(r)dt^{2}+ \frac{1}{N(r)}dr^{2}+r^{2}d\Omega^{2},
\end{equation}
where $N(r)$ is the (negative) constant curvature solution of $F(R)$-Maxwell theory \cite{moon}

\begin{equation}\label{eq11}
  N(r)=1-\frac{2m}{r}+\frac{Q^{2}}{(1+f_{R}(R_{0}))r^{2}}-\frac{R_{0}}{12}r^{2},
\end{equation}
where $m$ and $Q$ \footnote{Given that the charge $Q$ is constrained by the positive definiteness of Hawking temperature $T_{H}=\frac{N'(r_{h})}{4\pi}=\frac{1}{4\pi}\left[\frac{1}{r_{h}}-\frac{Q^2}{(1+f_{R}(R_{0}))r_{h}^3}-
\frac{R_{0}r_{h}}{4}\right]$, then $Q\leq \sqrt{(1+f_{R}(R_{0}))(r_{h}^2+\frac{3r_{h}^4}{L^2})}\equiv Q_{c}$, where $Q_{c}$ is the critical charge \cite{moon, nami}. For a small black hole $\frac{r_{h}}{L} \ll 1$ and the critical charge becomes $Q_{c}=\sqrt{(1+f_{R}(R_{0}))}r_{h}(1+\frac{3r_{h}^2}{2L^2})\ll L$.} are the mass and charge of the black hole and $f_{R}$ indicating derivative with respect to $R$.

\subsection{Superradiant instability}

Since the constant negative curvature solutions of $F(R)$ theory is the same as that of AdS solutions in GR provided that $R_{0} = -\frac{12}{ L^2} = 4\Lambda$ \cite{hendi, cruze, she}, we do not expect to derive new superradiant conditions because we deal with the same Klein-Gordon equation. However, a review of such conditions for a test scalar field scattered off a charged black hole \cite{pani} would be in order at this stage.

To study superradiance of a scattered test field off a charged black hole represented by (\ref{eq10}), we conventionally suggest the following ansatz
\begin{equation}\label{eq500}
 \Phi (r,t,\theta,\phi)=\int d\omega \sum_{lm}e^{-i\omega t}Y_{lm}(\theta ,\phi)\frac{\phi(r)}{r},
\end{equation}
where $Y_{lm}(\theta,\phi)$ are the usual spherical harmonics. We now fix the gauge potential according to $A_{\mu}=(-\frac{Q}{r}+C,0,0,0)$, where $C$ is an integration constant which is due to the requirement of having an electrically charged black hole with $A_{\mu}(r_{h})=0$ in the context of AdS/CFT, with the event horizon $r_{h}$ defined as the largest root of $N(r_{h})=0$, and is given by $C=\frac{Q}{r_{h}}$ \cite{nami}. It is interesting to note that in a gauge for which $\phi$ is real, the current is given by $\phi^2 A_{\mu}$  and must be finite at the event horizon, so that we need to have $\phi(r_{h})^2 A_{\mu}(r_{h})=0$. Then the assumption $\phi(r_{h})=0$ leads to a constant scalar field and we must consider $A_{\mu}(r_{h})=A_{t}(r_{h})=0$ \cite{horowitz}.

Substituting the above ansatz into the Klein-Gordon equation (\ref{eq55}), results in the following Schrodinger-like equation
\begin{equation}\label{eq5}
\frac{d^{2}\phi}{dr_{*}^{2}}+V_{eff} \phi=0,
\end{equation}
where use has been made of tortoise coordinate $r_{*}$, defined according to $r_{*}\equiv\int\frac{dr}{N(r)}$, with the effective potential given by
\begin{equation}\label{eq13}
V_{eff}=-N\left(\frac{l(l+1)}{r^{2}}+\frac{N'}{r}\right)+\left(\omega-qQ(\frac{1}{r}-\frac{1}{r_{h}})\right)^2.
\end{equation}
Using the asymptotic behavior of the effective potential above, one can derive the radial solutions near horizon ($r_{*}\rightarrow -\infty$) and at infinity ($r_{*}\rightarrow +\infty$) \cite{nami}
\begin{eqnarray}\label{14}
  &&\phi \sim \frac{A}{r^2}                         \qquad r_{*}\rightarrow +\infty ,\nonumber\\
  &&\phi \sim e^{-i\omega r_{*}}                     \qquad r_{*}\rightarrow -\infty,
\end{eqnarray}
where $A$ is a constant. By matching the asymptotic expansion of (\ref{14}), Uchikata and Yoshida derived the instability condition for the RN-AdS black hole \cite{nami}
\begin{equation}\label{eq7}
\mbox{Re}(\omega)<0.
\end{equation}
Note that this result is valid only for small black holes, $r_{h}\ll L$. As argued in \cite{nami}, this result confirms that the normal modes satisfy the superradiant condition $\mbox{Re}(\tilde{\omega})r_{h}-qQ<0$ \footnote{Due to the behavior of the radial function near the horizon (\ref{14}), an observer far from the black hole could see that the waves are coming out
from the black hole provided that $\omega$ satisfies superradiant condition (\ref{eq7}). In contrast, due to the negative group velocity, a local observer of the black hole sees only waves going into the horizon  and therefore the boundary condition is satisfied \cite{nami}.}, where $\mbox{Re}(\tilde{\omega})=\mbox{Re}(\omega)+\frac{q Q}{r_{h}}$ shows a shift of the real part of the frequency \footnote{Note that the superradiant condition is given by $ \omega r_{h}-q Q< 0$ for the $RN$ black hole.}.

\begin{figure}[!ht]
\includegraphics[width=8.2cm,height=4.5cm]{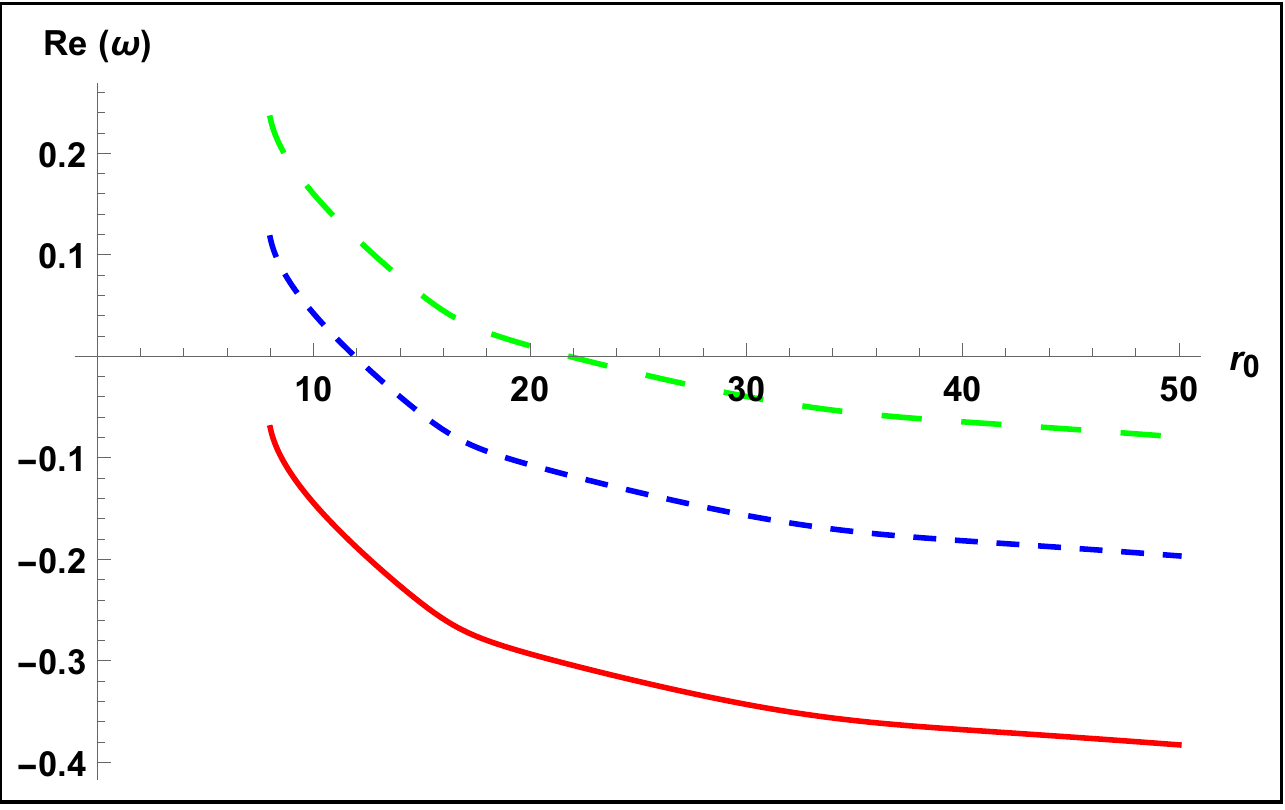}
\includegraphics[width=8.2cm,height=4.5cm]{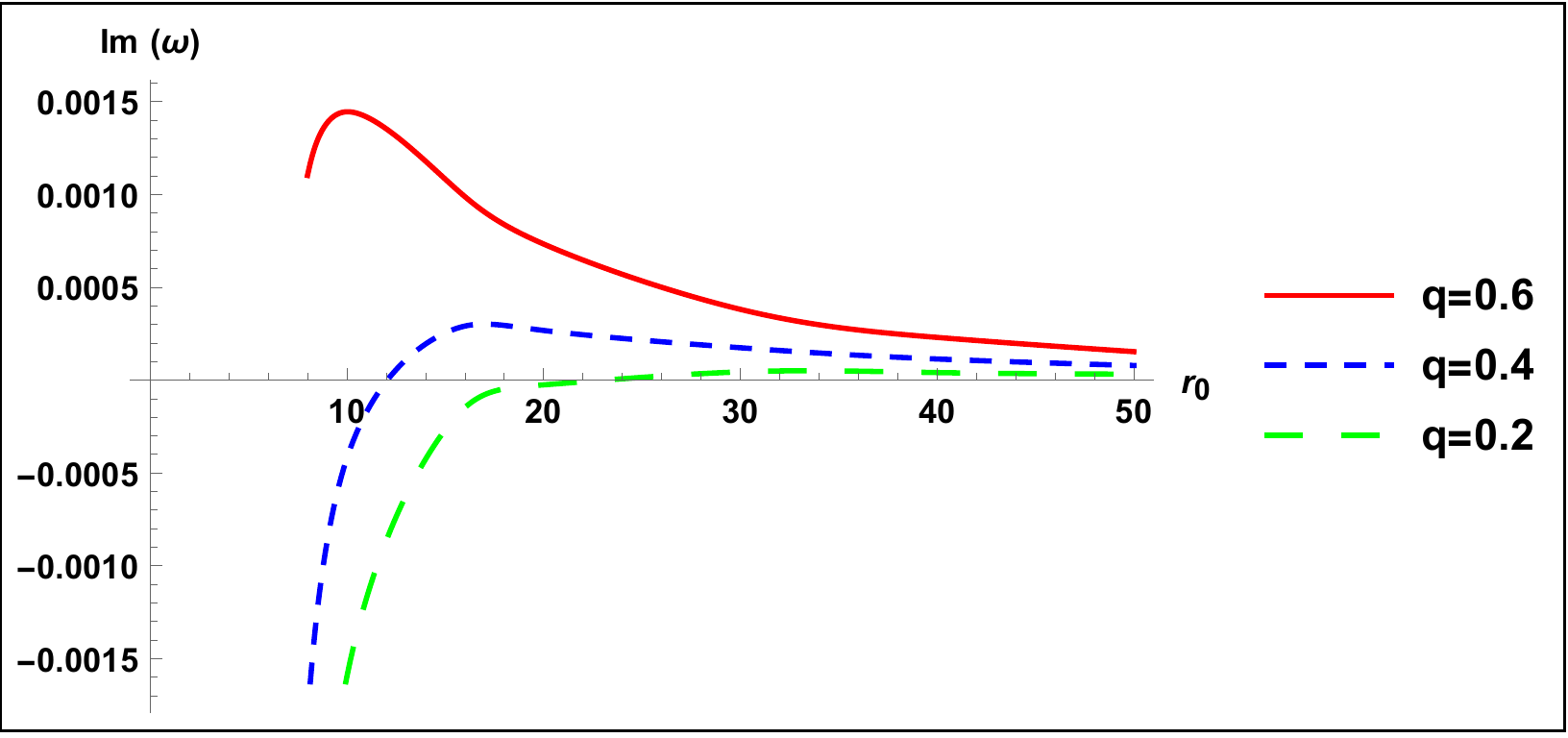}\\
\includegraphics[width=8.2cm,height=4.5cm]{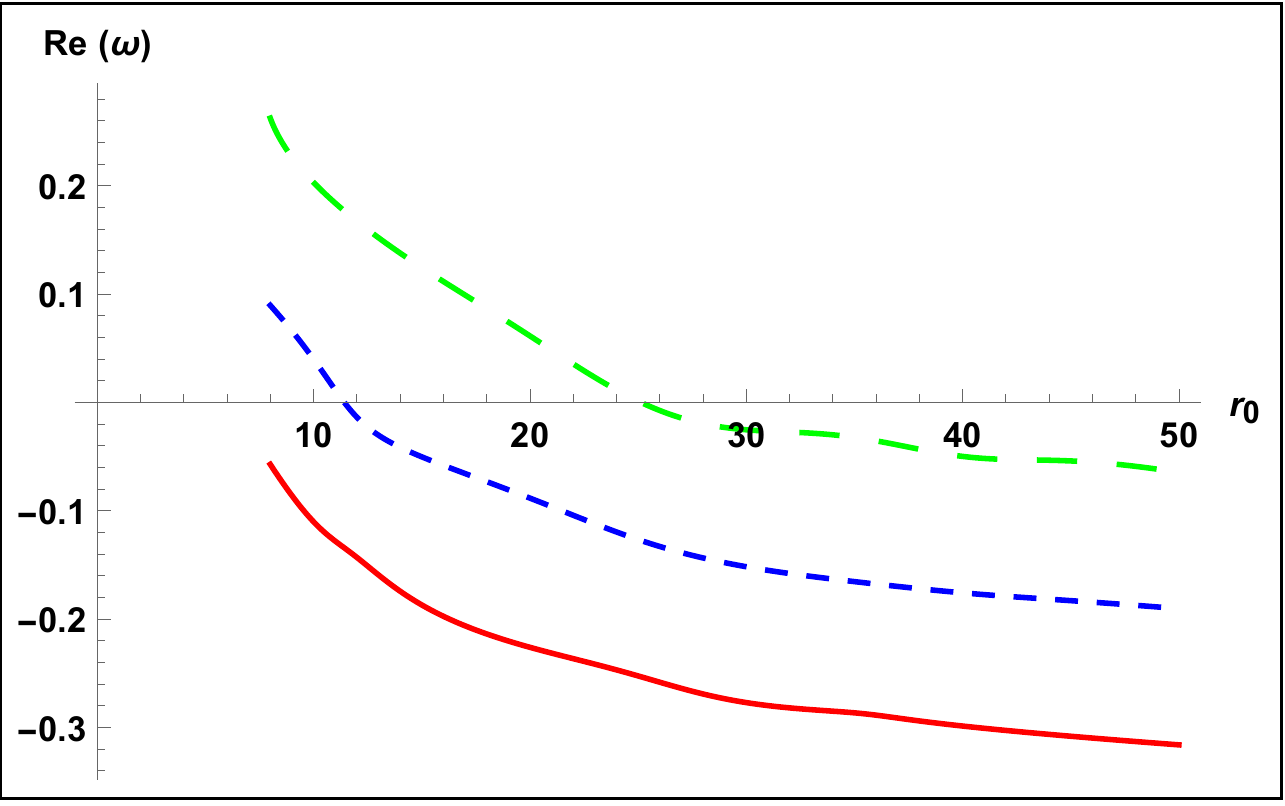}
\includegraphics[width=8.2cm,height=4.5cm]{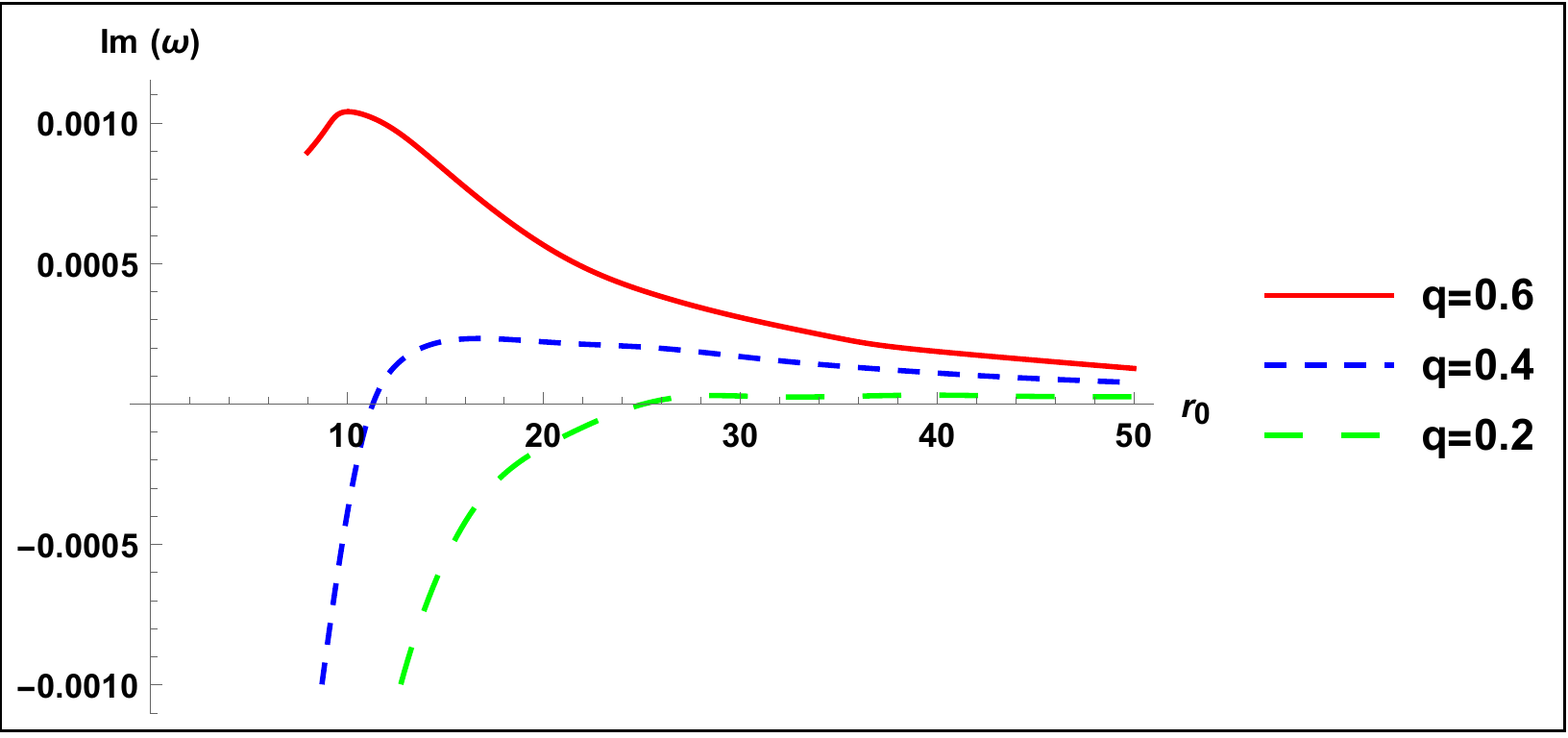}
\caption{\footnotesize{The real and imaginary parts of the frequency as a function of the AdS radius and different values of $q$ with $m=1$, $Q=0.9$. Top:  $F(R)=R-\frac{\mu^{4}}{R}$. Bottom: $F(R)=R-\lambda e^{-\alpha R}$ with $\alpha=1$. Note that $\mu^{2}$, $\lambda$ are obtained by fixing $R_{0}$ (see Appendix A ).}}
\label{ffig7}
\end{figure}

\begin{figure}[!ht]
\includegraphics[width=8.2cm,height=4.5cm]{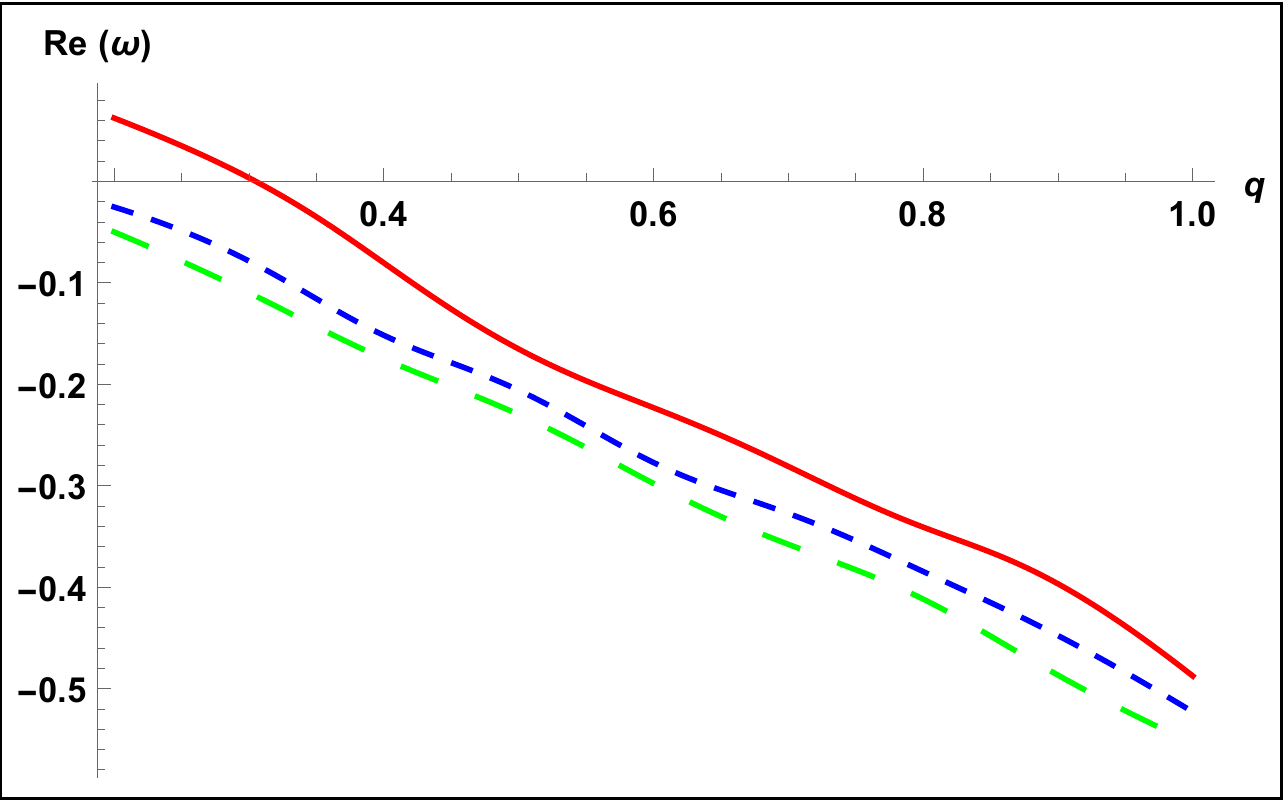}
\includegraphics[width=8.2cm,height=4.5cm]{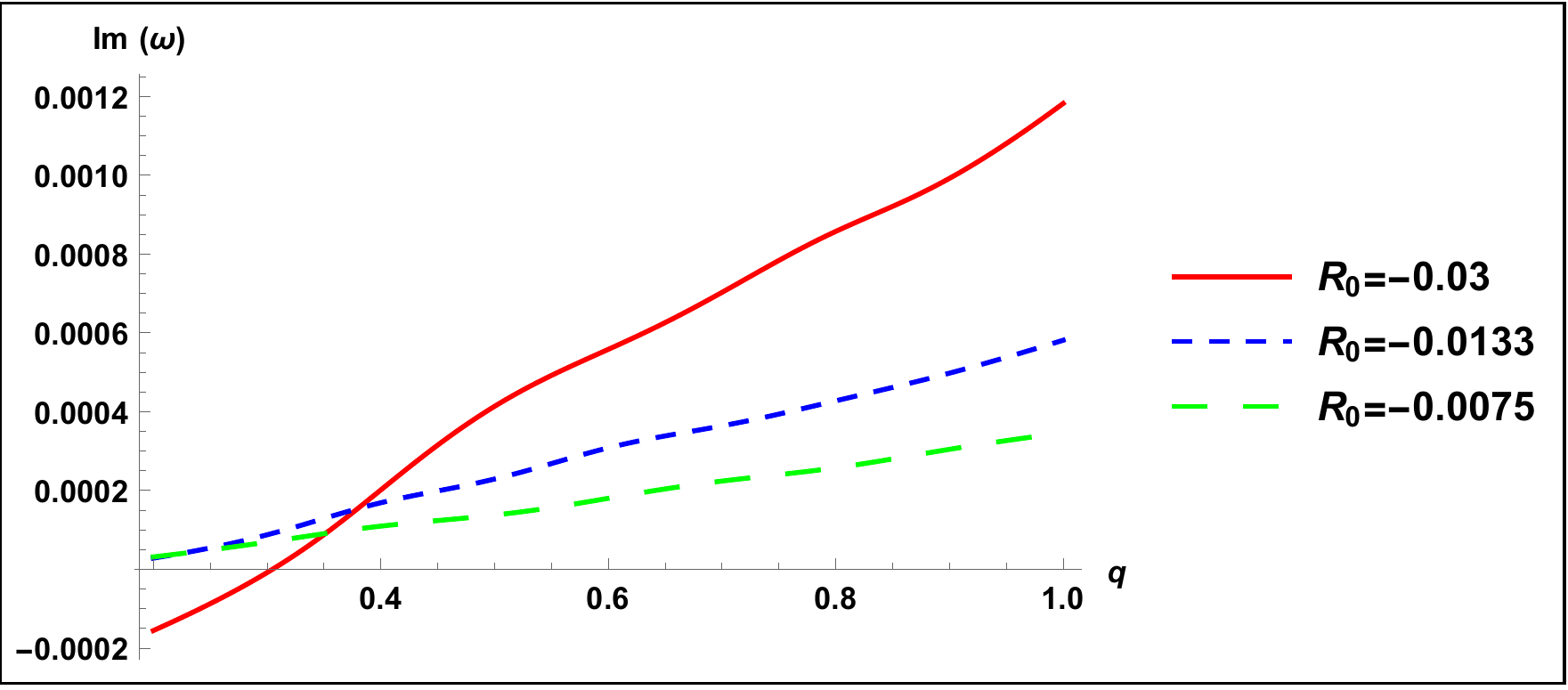}
\caption{\footnotesize{The real and imaginary parts of the frequency as a function of $q$ and different values of the AdS radius for $F(R)=R-\lambda e^{-\alpha R}$ with $m=1$, $Q=0.9$ and $\alpha=1$.}}
\label{fffig7}
\end{figure}

Summarising, based on the results of \cite{nami}, one may conclude that small Reissner-Nordstrom AdS black holes are unstable against charged scalar perturbations due to \emph{superradiance}. Furthermore, the authors show analytically that the real and imaginary parts of the frequency for the lowest order modes are
\begin{eqnarray}\label{36}
&&\mbox{Re}(\omega)= \frac{3}{L}-\frac{q Q}{r_{h}},\nonumber\\
&&\mbox{Im} (\omega)= -\emph{c} \frac{\mbox{Re} (\omega)}{L^2},
\end{eqnarray}
where $\emph{c}=\frac{16}{3\pi} r_{h}^2$ \cite{nami}.
Thus, it is clear that while (\ref{eq7}) is satisfied, the imaginary part of the frequency will be positive which means that the wave function grows exponentially and the black hole is unstable. It is also well-known that the AdS boundary condition is that of the Dirichlet type which ensures that there is no energy dissipation at the asymptotic boundary and therefore can be specified as a reflective boundary \cite{oscar}
\begin{equation}\label{eq203}
\phi(r_{0}=L)=0,
\end{equation}
In other words, an AdS space-time black hole can be modeled as a black hole in a closed system with a mirror-like boundary \cite{nami}. Practically, to investigate the instability for any $F(R)$ model, one repeatedly  integrates equation (\ref{eq5}) numerically, from $r_{h}$ to $r_{0}$, the AdS radius, while changing the fundamental frequency (\ref{36}) until $\phi(r_{0})=0$ is achieved, see Figs. \ref{ffig7} and \ref{fffig7} for $F(R)=R-\frac{\mu^{4}}{R}$ and $F(R)=R-\lambda e^{-\alpha R}$ \cite{huang}.

In Fig. \ref{ffig7}, the onset of instability ($r_{c}$) is where the imaginary part of the frequency becomes zero, the threshold frequency. The real part of  the frequency at $r_{c}$ becomes zero too and Figs. \ref{ffig7} and \ref{fffig7} show the sign change of  $\mbox{Re} (\omega)$ at the threshold frequency where the scalar field experiences superradiant instability in the negative areas. Given Fig. \ref{ffig7} and taking the AdS boundary as the enclosing boundary, if the AdS radius is small, $r_{0}<r_{c}$, then $\mbox{Im} (\omega)$ is negative and the modes decay exponentially. If however the AdS radius is large, $r_{0}>r_{c}$, then $\mbox{Im}(\omega)$ is positive and the modes grow exponentially which leads to superradiant instability. Furthermore the growth of superradiant modes will occur at large AdS radii for small $q$ values.

\section{Static hairy black hole solutions}
Due to the existence of superradiant instability in charged $F(R)$  black holes under certain conditions, we first seek static hairy black hole solutions and continue to investigate their stability to confirm that these solutions can be regarded as the possible endpoint of superradiant instability. To get static solutions, let us  remove the time dependence of $\Phi$ with the use of gauge freedom $\Phi\rightarrow e^{i \chi}\Phi$ by assuming $\chi=\omega t$. Then the scalar field depends on $r$ ,$\Phi=\phi(r)$, and the gauge potential is $A_{\mu}=(A_{0}(r),0,0,0)$ which may enable us to obtain static solutions with a nontrivial scalar hair in a $F(R)$ theory near the threshold frequency, $\omega=\omega_{c}$\footnote{We construct hairy black hole solutions near the critical frequency ($\omega_{c}$) which is the onset of superradiant instability \cite{oscar}, since it seems sensible to expect superradiant instability to end as a hairy black hole.}. We consider a static, spherically symmetric space-time as
\begin{eqnarray}\label{17}
ds^{2} =-N(r)h(r)dt^{2}+ \frac{1}{N(r)}dr^{2}+r^{2}d\Omega^{2},
\end{eqnarray}
where
\begin{eqnarray}\label{16}
N(r)=1-\frac{2 m(r)}{r}+\frac{ r^2}{L^2},
\end{eqnarray}
and the metric variable $h(r)$ shows the back reaction of the scalar field on the background \cite{peng}. Inserting (\ref{17}) into field equations (\ref{eq3}-\ref{6}), one finds
\begin{equation}\label{18}
(1+f_{R}) h'(r)=\frac{r q^2 A_{0}(r)^2\phi(r)^2}{N(r)^2}+r h(r) \phi'(r)^2,
\end{equation}
\begin{equation}\label{19}
A'_{0}(r)^2=-\frac{2(1+f_{R})}{r}\left[h(r)\left(N'(r)+\frac{N(r)}{r}-\frac{1}{r}\right)+\frac{1}{2} N(r) h'(r)\right] -f_{R} h(r) R_{0}+f h(r),
\end{equation}
\begin{equation}\label{20}
N(r)A''_{0}(r)+\left(\frac{2N(r)}{r}-\frac{N(r) h'(r)}{2 h(r)}\right)A'_{0}(r)-q^2\phi(r)^2 A_{0}(r)=0,
\end{equation}
\begin{equation}\label{21}
N(r)\phi''(r)+\left(\frac{2N(r)}{r}+N'(r)+\frac{N(r)h'(r)}{2h(r)}\right)\phi'(r)+\frac{q^2 A_{0}(r)^2}{N(r)h(r)}\phi(r)=0.
\end{equation}
Here a prime shows partial derivatives with respect to $r$. In the rest of the paper, we will refer to (\ref{18}-\ref{21}) as the equations of motion. In the absence of any analytical solution, we integrate them numerically.
\subsection{Boundary conditions}
Equations (\ref{18}-\ref{21}) cannot be solved analytically for a non-zero scalar field. To solve them numerically we need boundary conditions at the horizon and AdS boundary. Assuming we have a regular event horizon defined as $N(r_{h})=0$, we find $m(r_{h})\equiv m_{h}=\frac{r_{h}}{2}(1+\frac{r_{h}^2}{L^2})$.
We also assume that the field variables are regular and analytic functions at the event horizon. Therefore Taylor expansions of the field variables in the neighborhood of the event horizon is given by
\begin{equation}\label{eq21}
m=m_{h}+m'_{h}(r-r_{h})+...,
\end{equation}
\begin{equation}\label{eq22}
h=h_{h}+h'_{h}(r-r_{h})+...,
\end{equation}
\begin{equation}\label{eq23}
A_{0}=A_{h}+A'_{h}(r-r_{h})+\frac{A''_{h}}{2}(r-r_{h})^2+...,
\end{equation}
\begin{equation}\label{eq24}
\phi=\phi_{h}+\phi'_{h}(r-r_{h})+\frac{\phi''_{h}}{2}(r-r_{h})^2+...,
\end{equation}
where $\phi_{h}\equiv \phi(r_{h}) $ and $A'_{h}\equiv A'(r_{h})$ which are the scalar and electric fields on the horizon respectively. Assuming $N'(r_{h})>0$ in equations (\ref{20}, \ref{21}) one has $ \phi'_{h}\equiv \phi'(r_{h}) =0$ and $ A_{h}\equiv A(r_{h})=0$, that is, the gauge vanishes at the horizon as expected.  Due to the freedom for choosing the coordinate system, we fix $h_{h}=1$ without loss of generality \cite{win}. By plugging relations (\ref{eq21}-\ref{eq24}) in equations (\ref{18}-\ref{21}), we obtain boundary conditions at the event horizon
\begin{eqnarray}\label{201}
&&m'_{h}=\frac{r_{h}^2}{4\left(1+f_{R_{0}}\right)}\left({A'_{h}}^2-f+f_{R_{0}}R_{0}\right)+\frac{3 r_{h}^2}{2 L^2},\nonumber \\
&&h'_{h}=\frac{4\left(1+f_{R_{0}}\right){r_{h}}^3 q^2 {\phi_{h}}^2{A'_{h}}^2}{\left[-{r_{h}}^2\left({A'_{h}}^2-f+f_{R_{0}}R_{0}\right)+2\left(1+f_{R_{0}}\right)\right]^2},\nonumber \\
&&A''_{h}=\frac{2q^2\phi_{h}^2{A'_{h}}\left(1+f_{R_{0}}\right)r_{h}}{-{r_{h}}^2\left({A'_{h}}^2-f+f_{R_{0}}R_{0}\right)+2\left(1+f_{R_{0}}\right)}-\left(\frac{2}{r_{h}}-\frac{h'_{h}}{2}\right){A'_{h}},\nonumber \\
&&\phi''_{h}=-\frac{2q^2 \phi_{h}{A'_{h}}^2\left(1+f_{R_{0}}\right)^2{r_{h}}^2}{\left[-{r_{h}}^2\left({A'_{h}}^2-f+f_{R_{0}}R_{0}\right)+2\left(1+f_{R_{0}}\right)\right]^2}.
\end{eqnarray}
It is worth noting that the AdS boundary condition is of the Dirichlet type which can be considered as a reflective boundary, confirming (\ref{eq203}). There are no constraints on variables $N(r)$, $h(r)$ and $A_{0}(r)$ at the AdS boundary \cite{win}.
\subsection{Numerical solution}\label{sub1}
We select the length scale by setting $r_{h}=1$ and rescale all quantities by $r_{h}$ to have dimensionless equations. With regard to selected models of $F(R)$, we integrate equations (\ref{18}-\ref{21}) numerically. We set $r=r_{h} +\epsilon$ with $\epsilon=10^{-8}$ since equations (\ref{18}-\ref{21}) are singular at $r=r_{h}$ and use the shooting method to replace initial conditions by boundary conditions at the event horizon.

\begin{figure}[!ht]
\includegraphics[width=8.2cm,height=4.5cm]{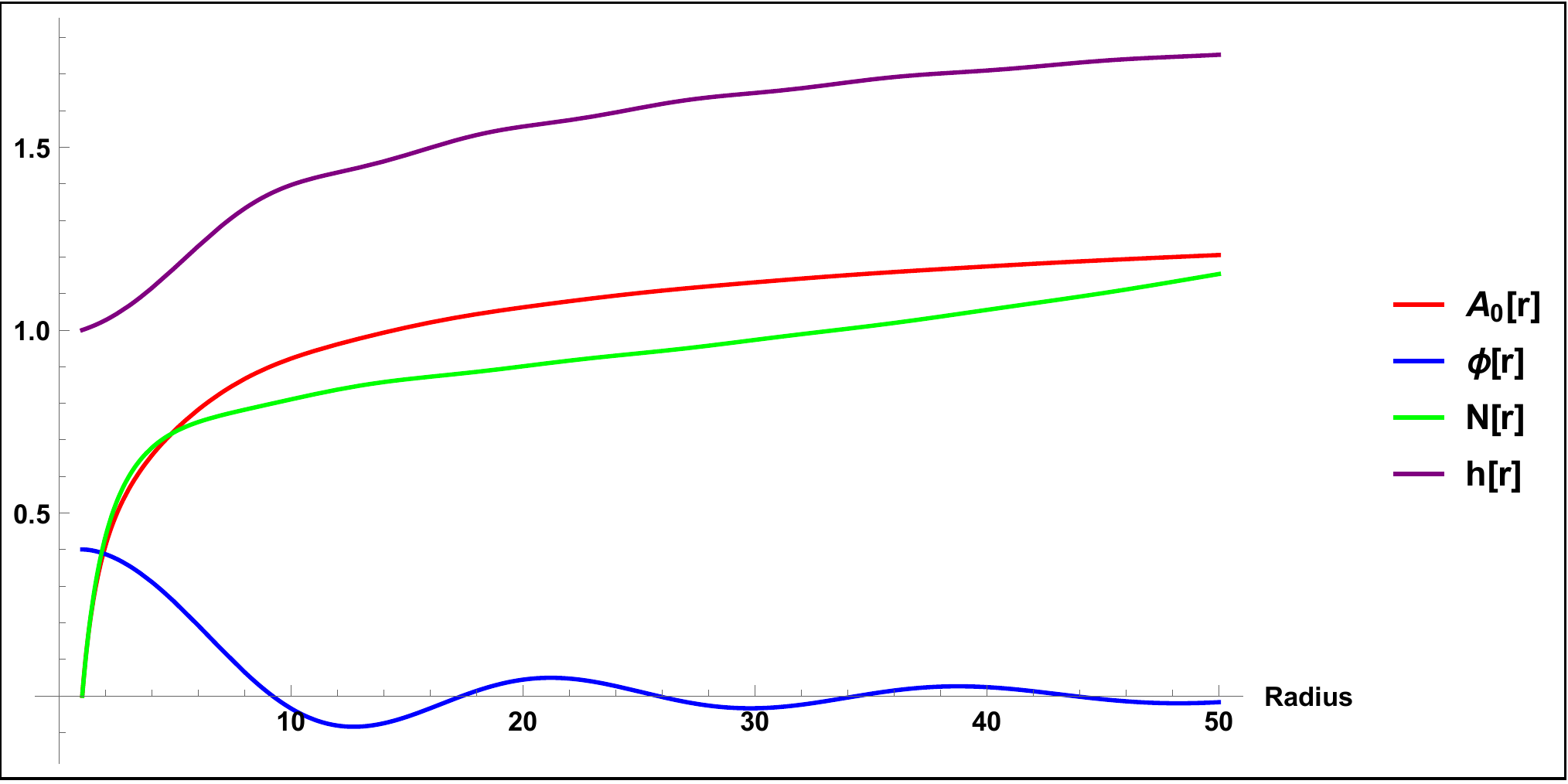}
\includegraphics[width=8.2cm,height=4.5cm]{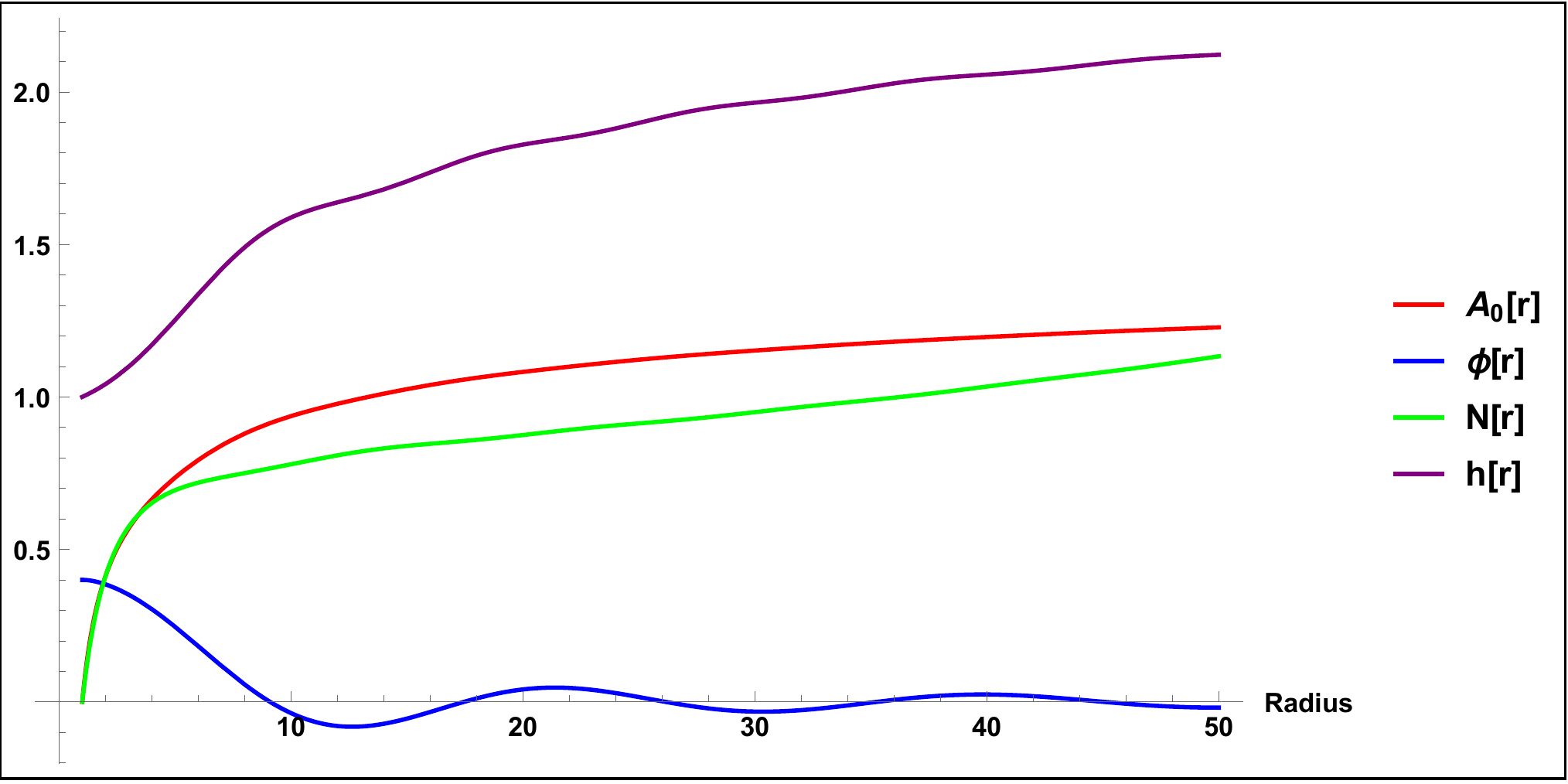}
\caption{\footnotesize{A plot of field variables as a function of radius with $q=0.4$, $A'_{h}=0.8$, $\phi_{h}=0.4$ and $R_{0}=-0.0012$ ($L=100$) for a static black hole. Left: $F(R)=R-\frac{\mu^{4}}{R}$. Right: $F(R)=R-\lambda e^{-\alpha R}$ with $\alpha=10$ where $\alpha$ is selected so that it satisfies the condition $-\frac{1}{\alpha}<R_{0}<0$, see Appendix A.}}\label{fig1}
\end{figure}

\begin{figure}[!ht]
\includegraphics[width=8.2cm,height=4.5cm]{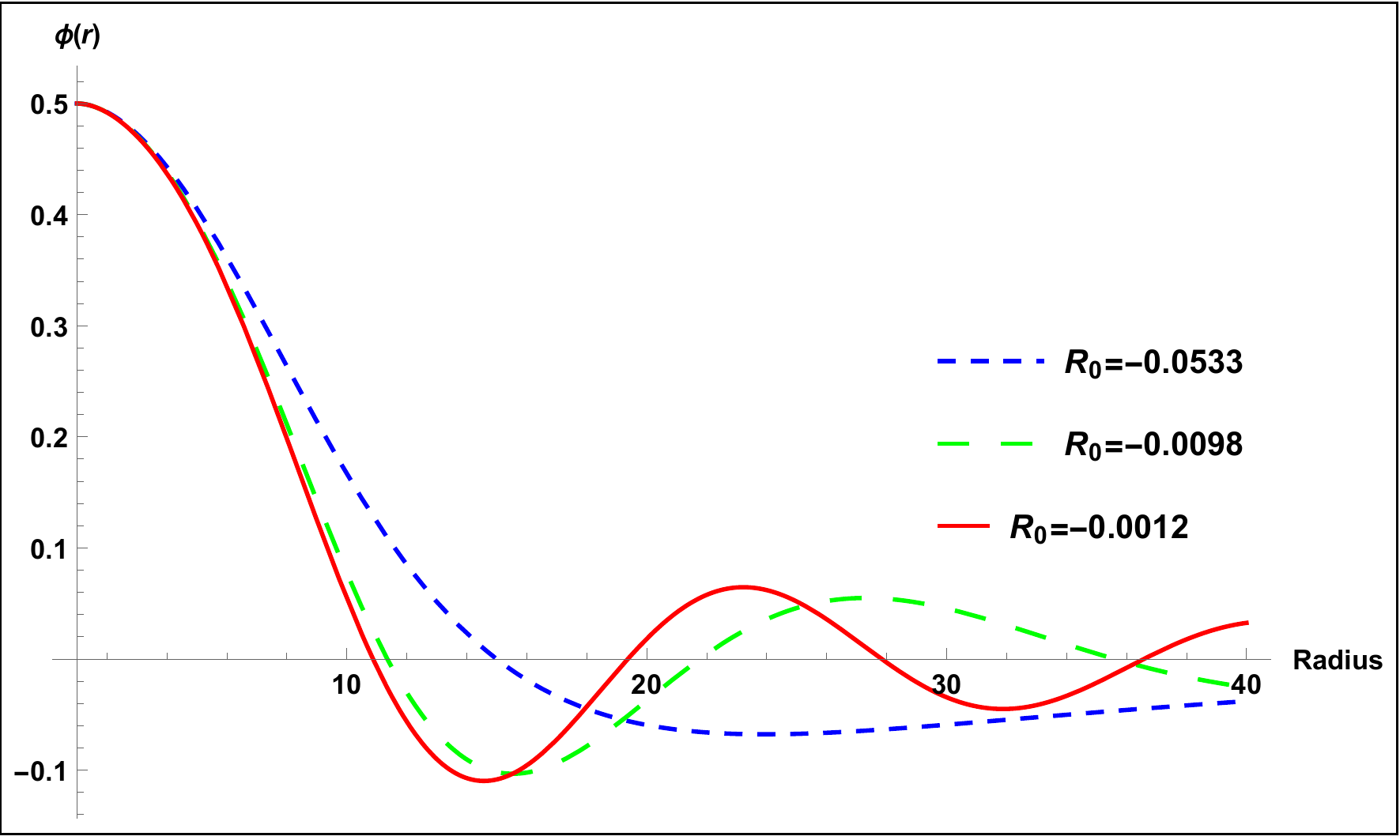}
\includegraphics[width=8.2cm,height=4.5cm]{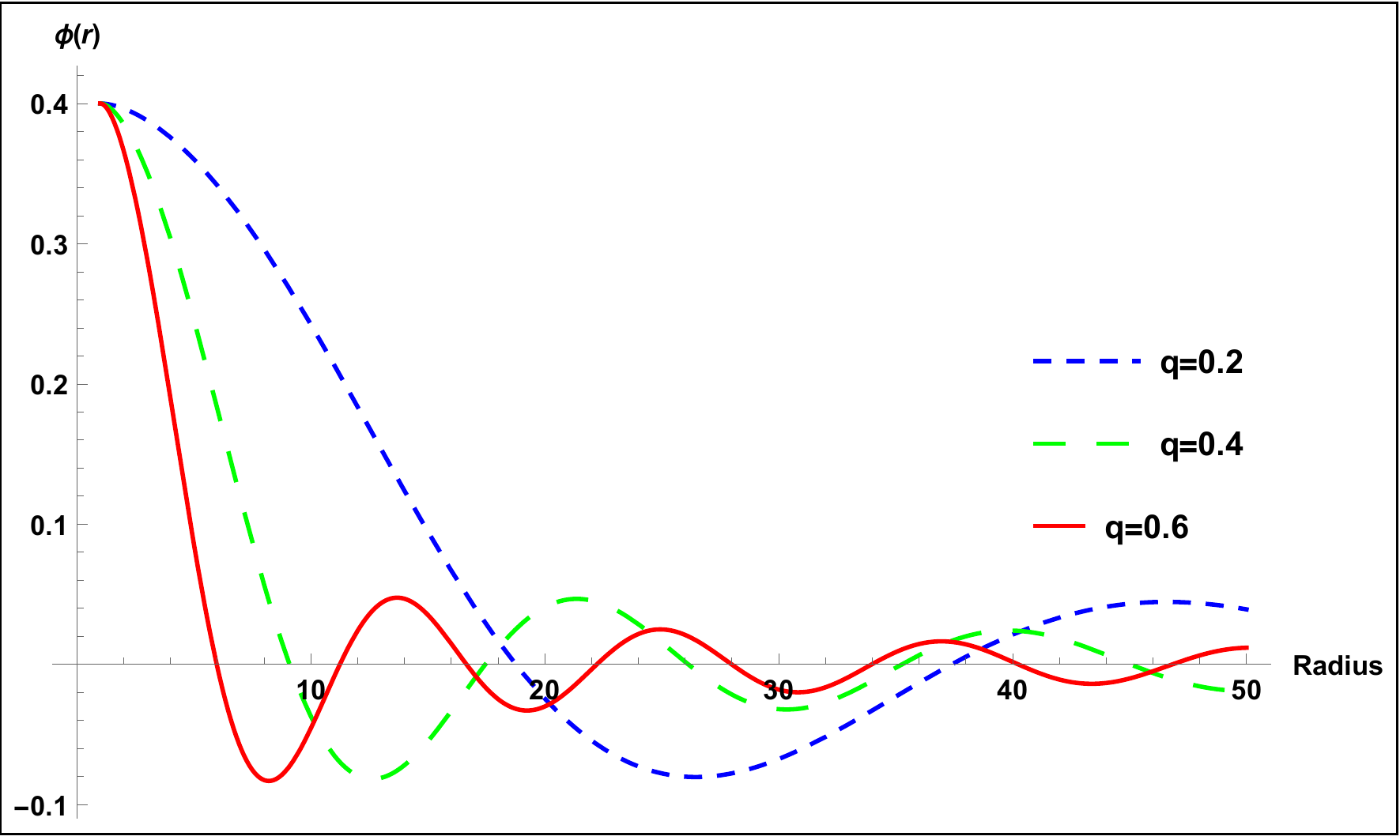}
 \caption{\footnotesize{ Three distinct scalar field profiles as a function of the radius. Left:  $F(R)=R-\frac{\mu^{4}}{R}$ with $q=0.4$, $\phi_{h}=0.5$, $A'_{h}=0.61$ and different values of $R_{0}$. Right:  $F(R)=R-\lambda e^{-\alpha R}$ with $\alpha=10$, $\phi_{h}=0.4$, $R_{0}=-0.0012$, $L=100$, $A'_{h}=0.8$ and different values of $q$.} }\label{fig2}
 \end{figure}

In Fig. \ref{fig1}, solutions of the field variables $N(r)$, $A_{0}(r)$, $h(r)$ and $\phi(r)$ are shown as a function of the radius for a spherically symmetric, static and asymptotically AdS black hole with $q=0.4$, $A'_{h}=0.8$, $\phi_{h}=0.4$ and $R_{0}=-0.0012$ for both models. The plots show the regular and nonsingular solutions outside the horizon. The scalar field outside the horizon fluctuates around zero and the field variables $N(r)$, $h(r)$ and $A_{0}(r)$  grow uniformly.
The left panel in Fig. \ref{fig2} shows different scalar field profiles outside the horizon with different AdS radii. The number of scalar field nodes increases by the growth of the AdS radius and for large AdS radii with the same initial conditions ($q$, $A'_{h}$ and $\phi_{h}$) the scalar field exhibits more radial nodes. The right panel in Fig. \ref{fig2} shows the increase in oscillations (nodes) of the scalar field as a function of charge.

\begin{figure}[!ht]
\includegraphics[width=8.2cm,height=4.5cm]{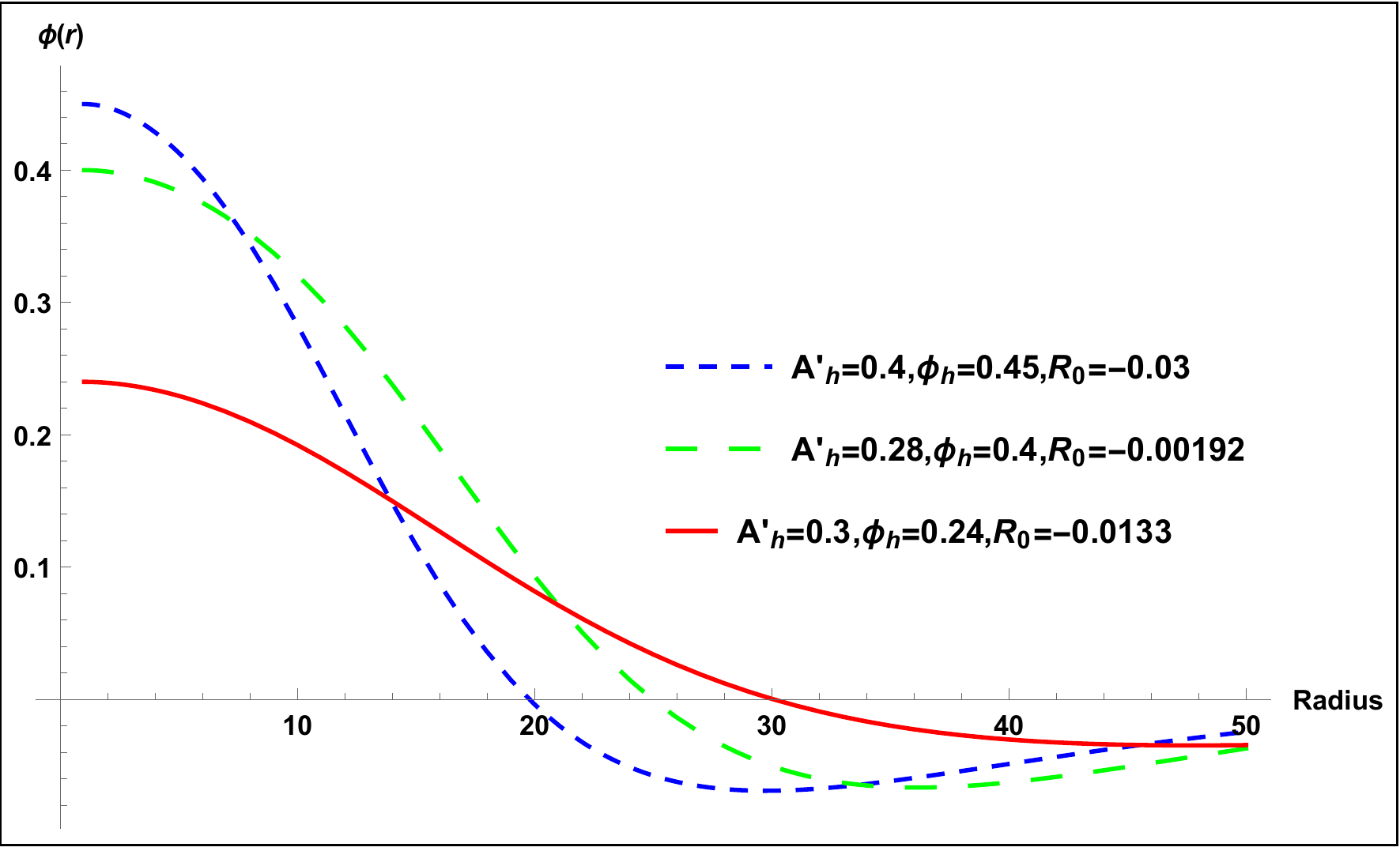}
\includegraphics[width=8.2cm,height=4.5cm]{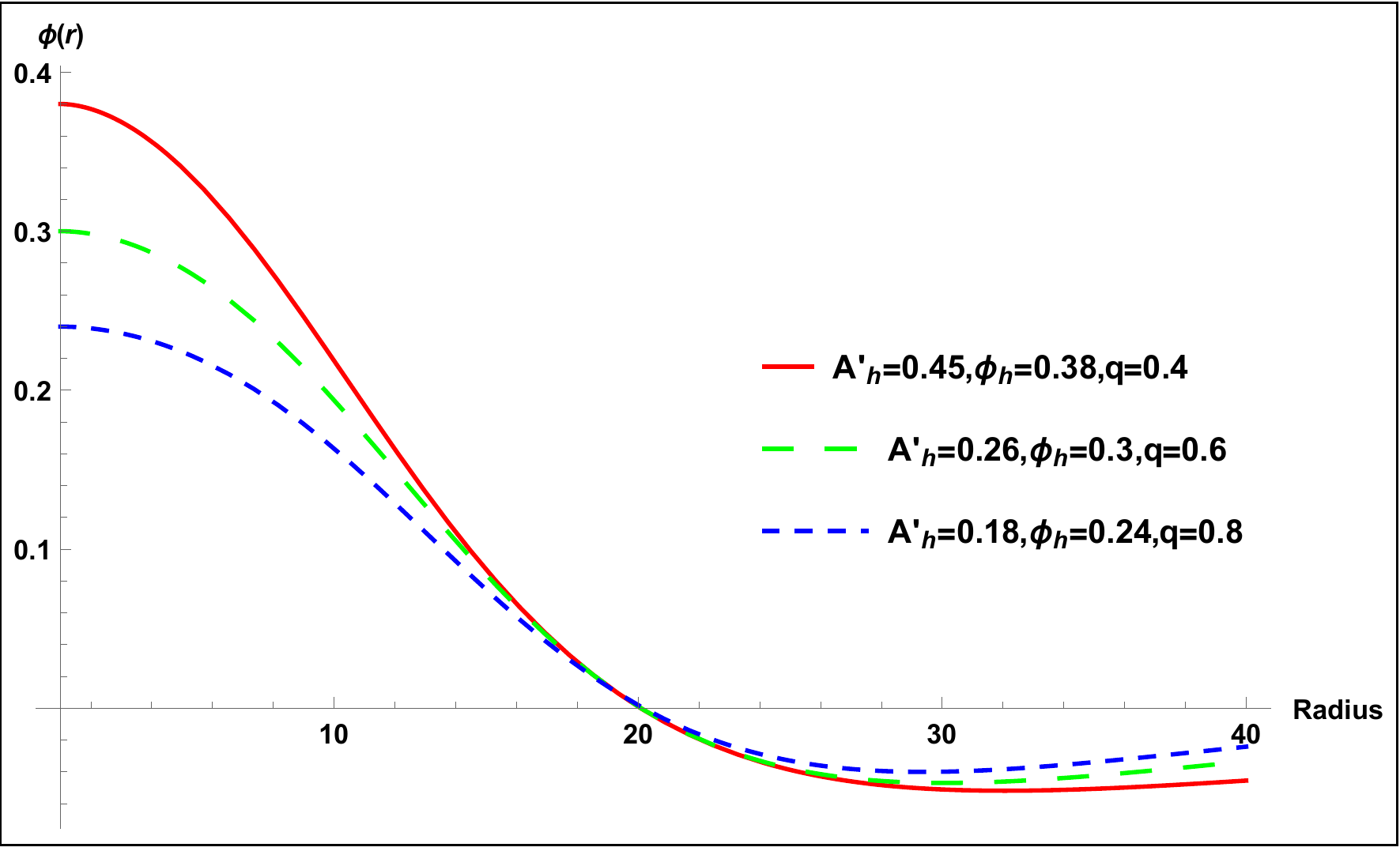}
\caption{\footnotesize{ Different scalar field profiles with different $A'_{h}$ and $\phi_{h}$ which have only one node at the AdS boundary. Left:  $F(R)=R-\frac{\mu^{4}}{R}$, $q=0.2$ and different values of the AdS radius. Right: $F(R)=R-\lambda e^{-\alpha R}$, $R_{0}=-0.03$ ($L=20$), $\alpha=10$ and different values of the scalar field charge. }}\label{fig2n}
\end{figure}

\begin{figure}[!ht]
\includegraphics[width=8.2cm,height=4.5cm]{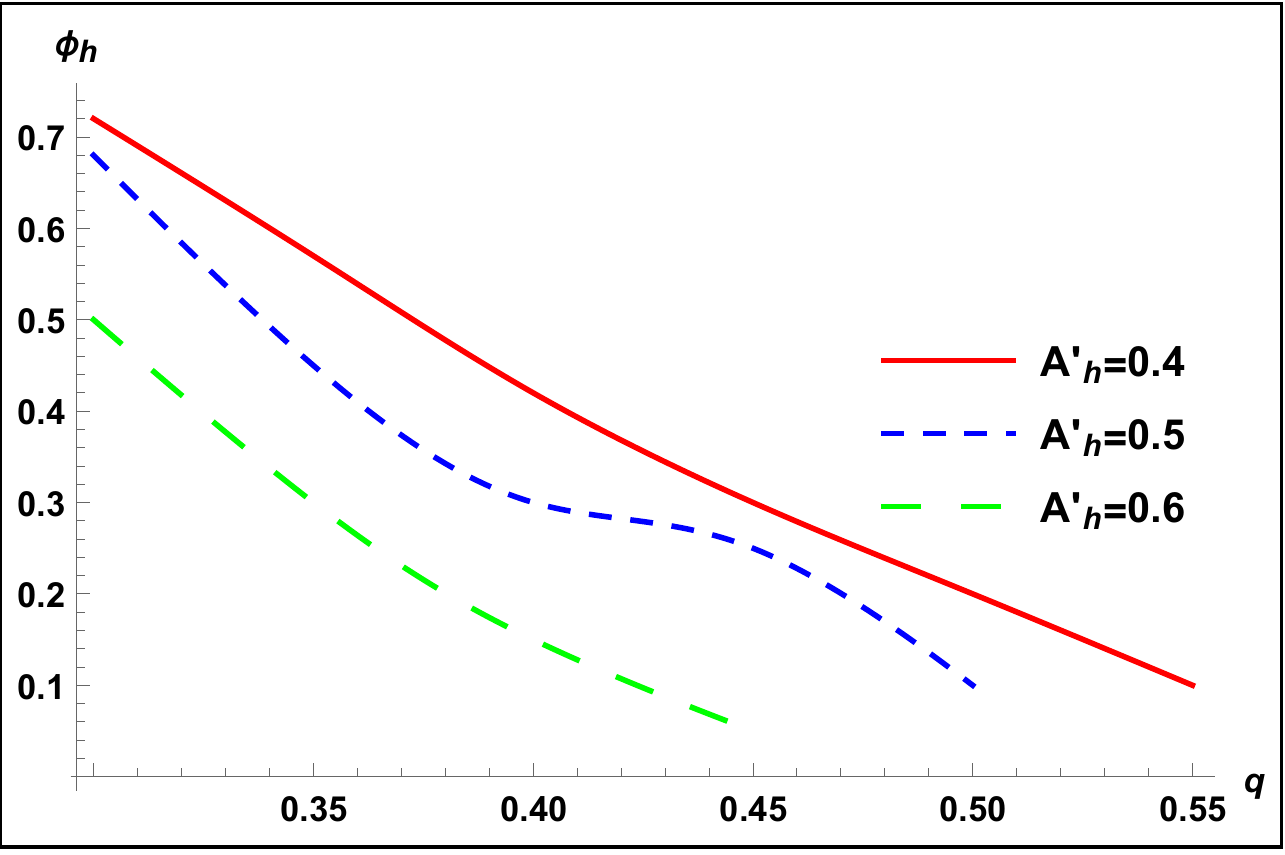}
\includegraphics[width=8.2cm,height=4.5cm]{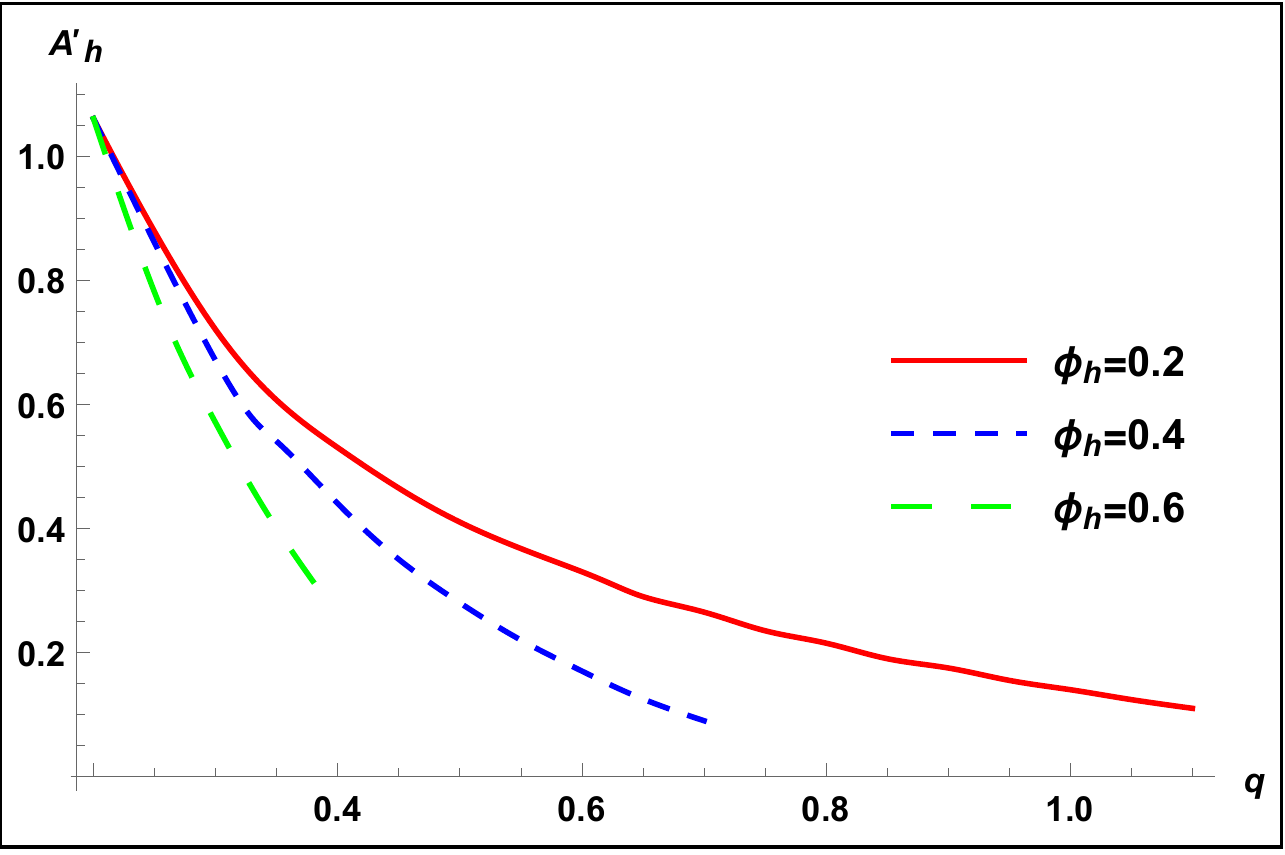}
\caption{\footnotesize{$\phi_{h}$ and $A'_{h}$ plotted as a function of $q$ when the scalar field has only one node at the AdS boundary $R_{0}=-0.03$ ($L=20$). Left: $F(R)=R-\frac{\mu^{4}}{R}$. Right: $F(R)=R-\lambda e^{-\alpha R}$ with $\alpha=10$.}}\label{fig20n}
\end{figure}

\begin{figure}[!ht]
\includegraphics[width=5.3cm,height=3cm]{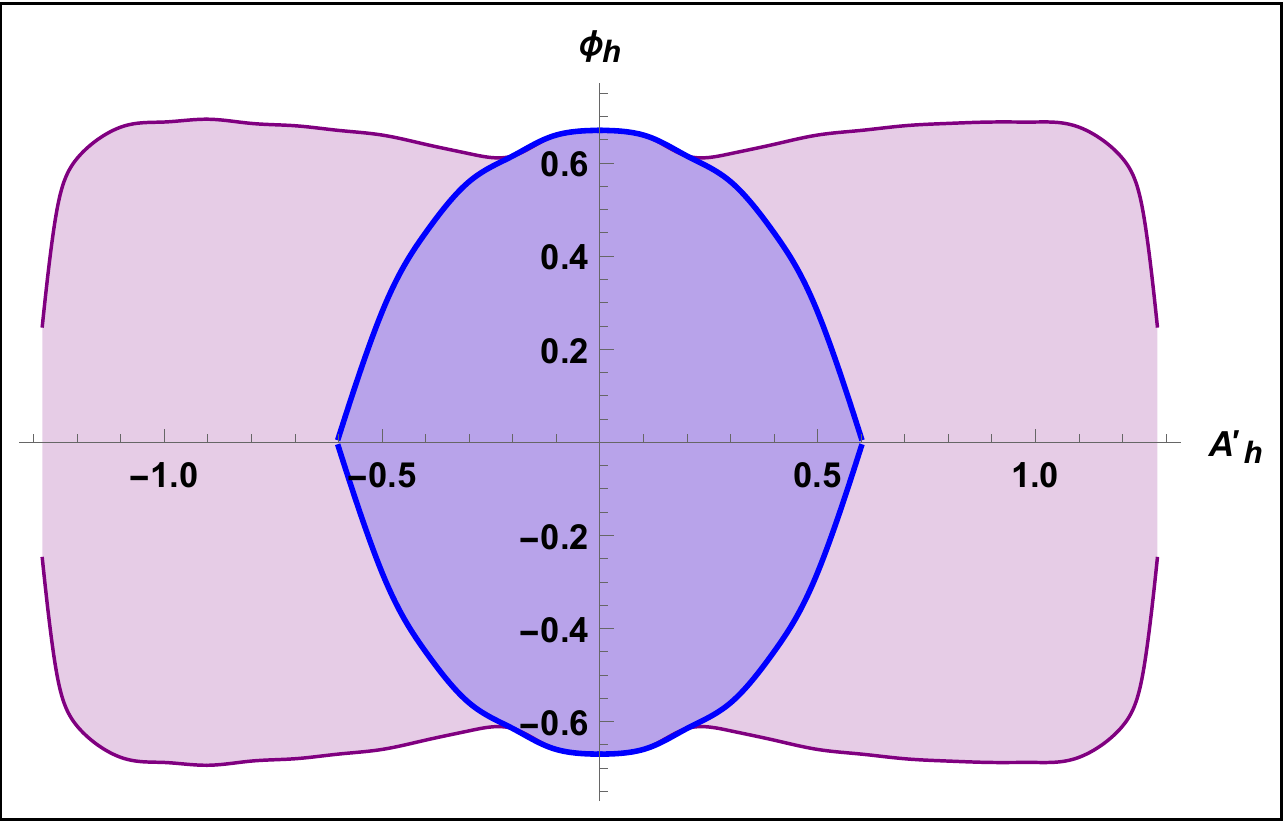}
\includegraphics[width=5.3cm,height=3cm]{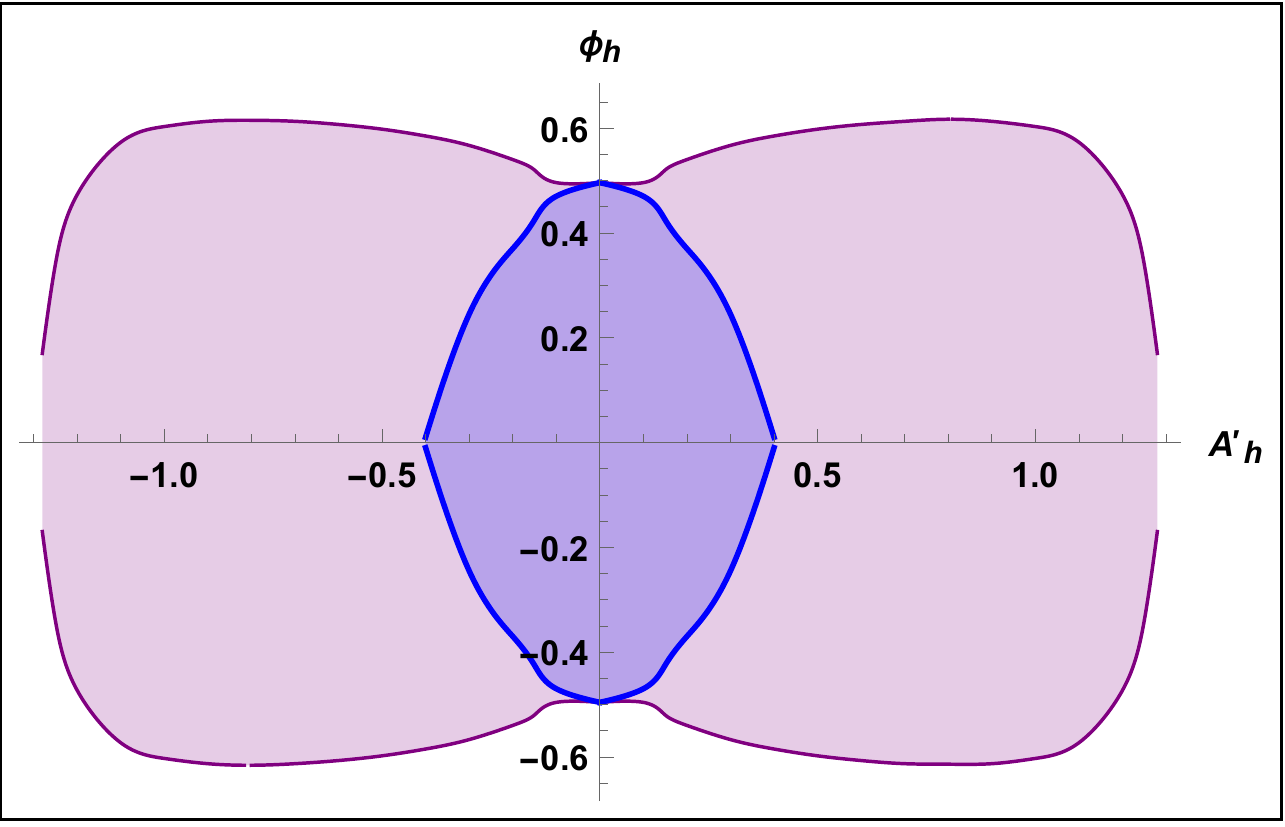}
\includegraphics[width=5.3cm,height=3cm]{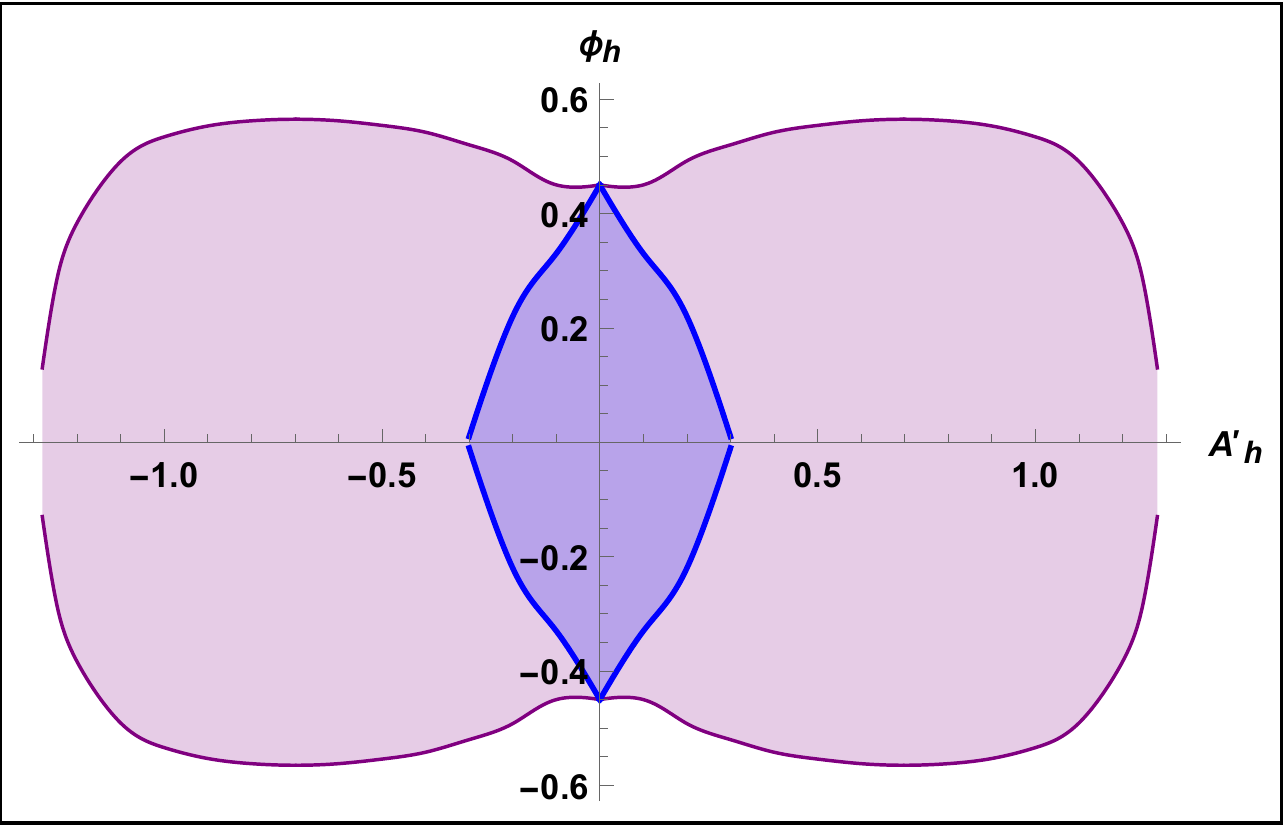}
\caption{\footnotesize{Phase space ($A'_{h}$-$\phi_{h}$) of black hole solutions for different values of $q=0.4,0.6,0.8$ with AdS radius $R_{0}=-0.03$ ($L=20$ ) and $\alpha=10$ for $F(R)=R-\lambda e^{-\alpha R}$. The purple regions and blue line show solutions with at least one node up to the AdS boundary. In the blue regions the scalar field has no node.}}\label{fig200n}
\end{figure}

\begin{figure}[!ht]
\includegraphics[width=5.3cm,height=3cm]{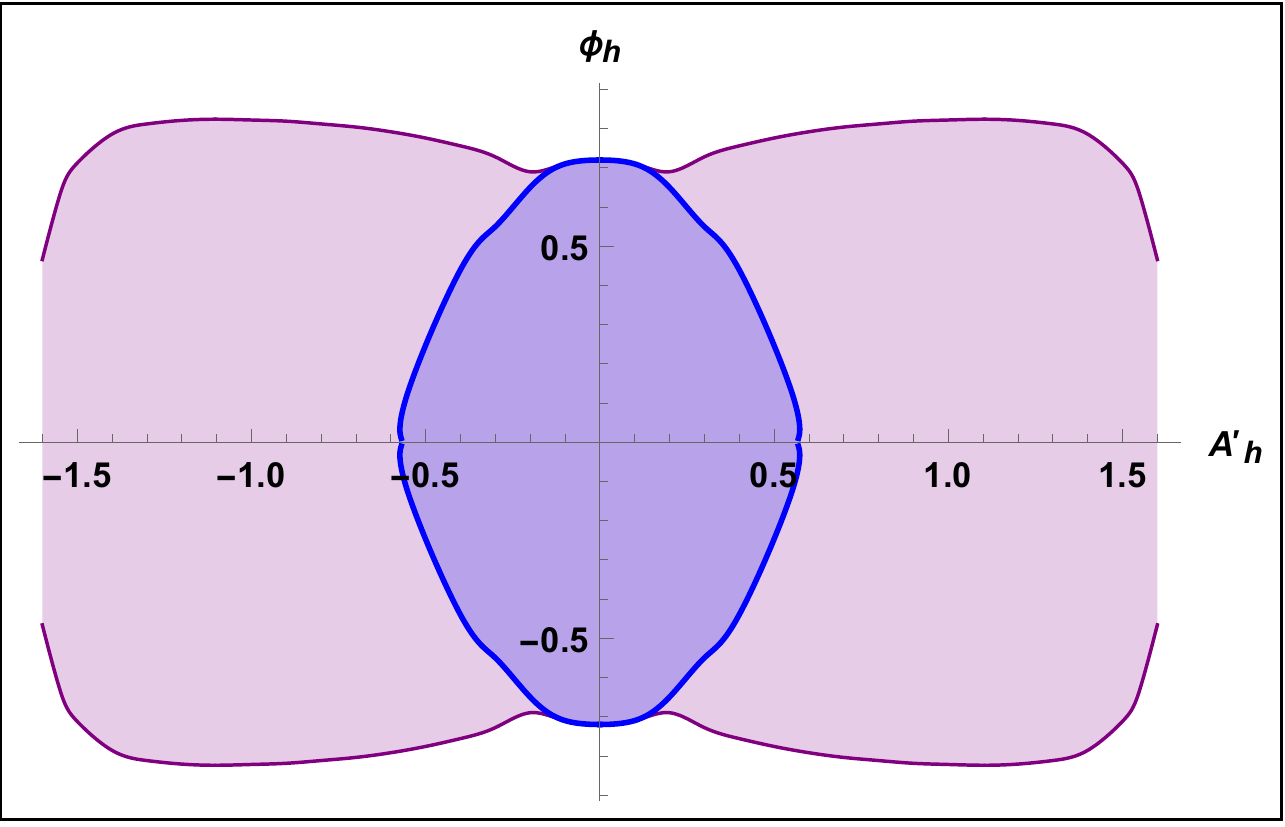}
\includegraphics[width=5.3cm,height=3cm]{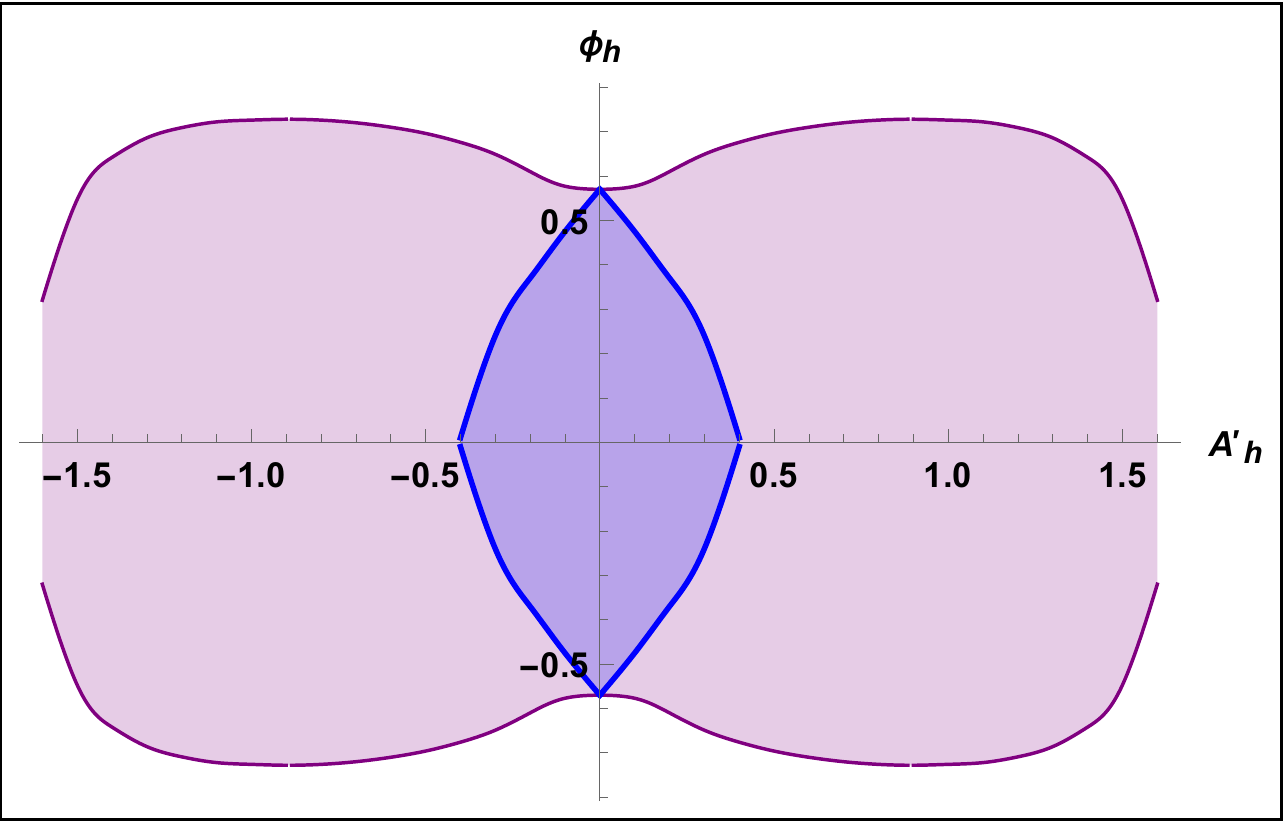}
\includegraphics[width=5.3cm,height=3cm]{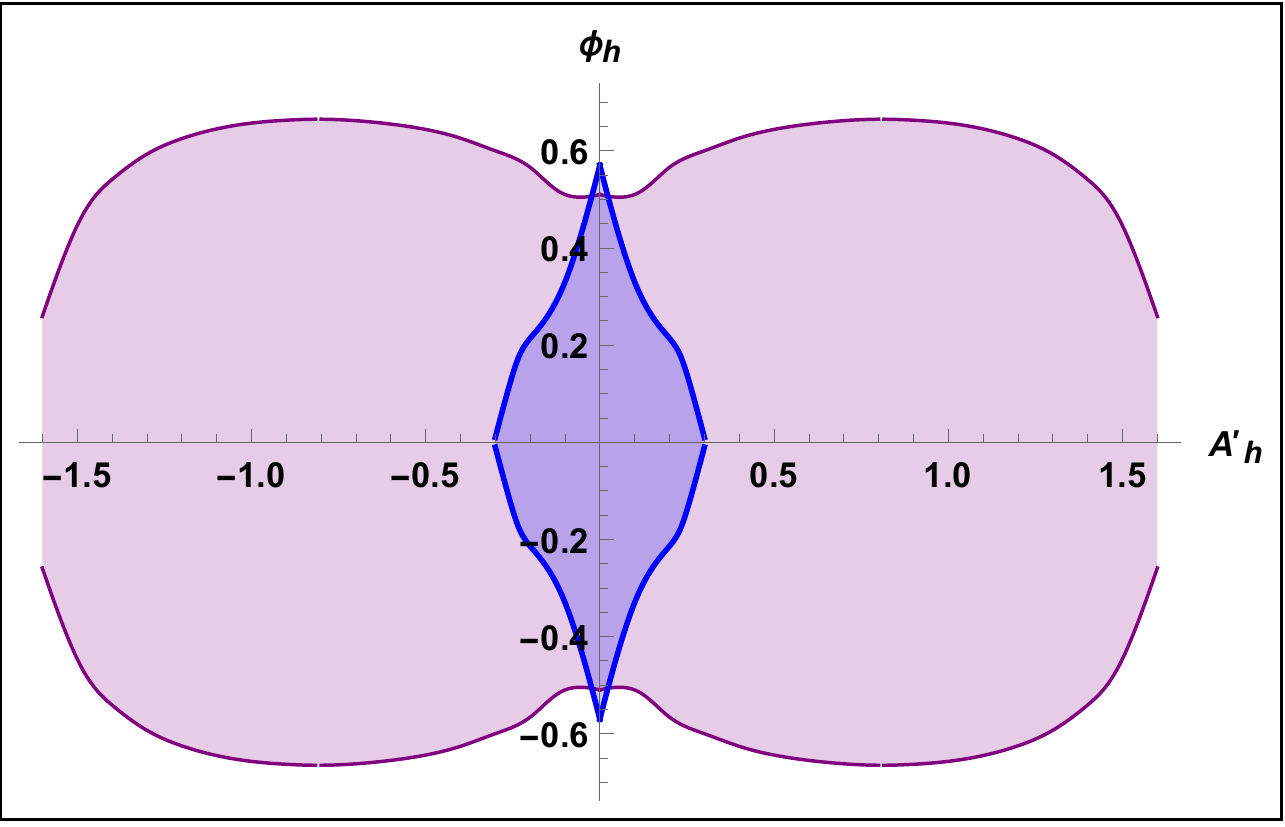}\\
\includegraphics[width=5.3cm,height=3cm]{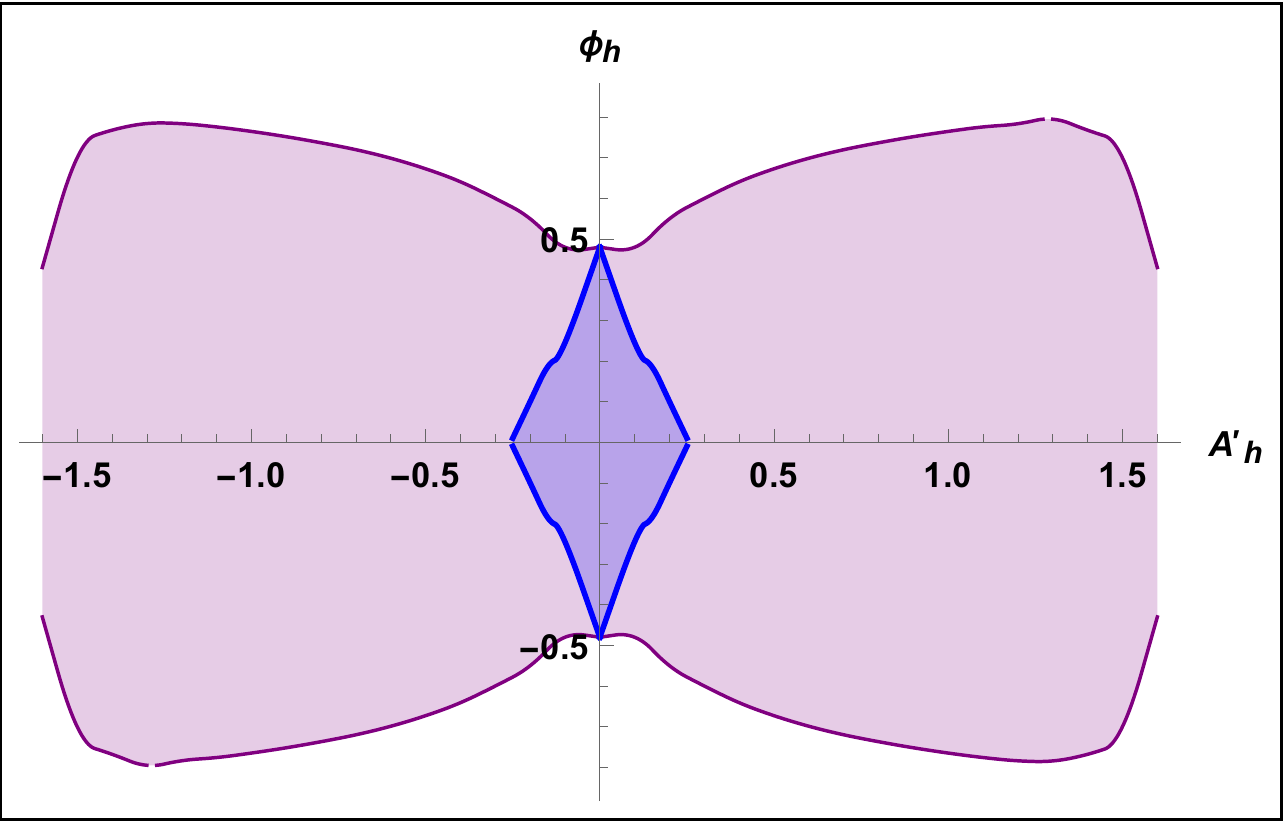}
\includegraphics[width=5.3cm,height=3cm]{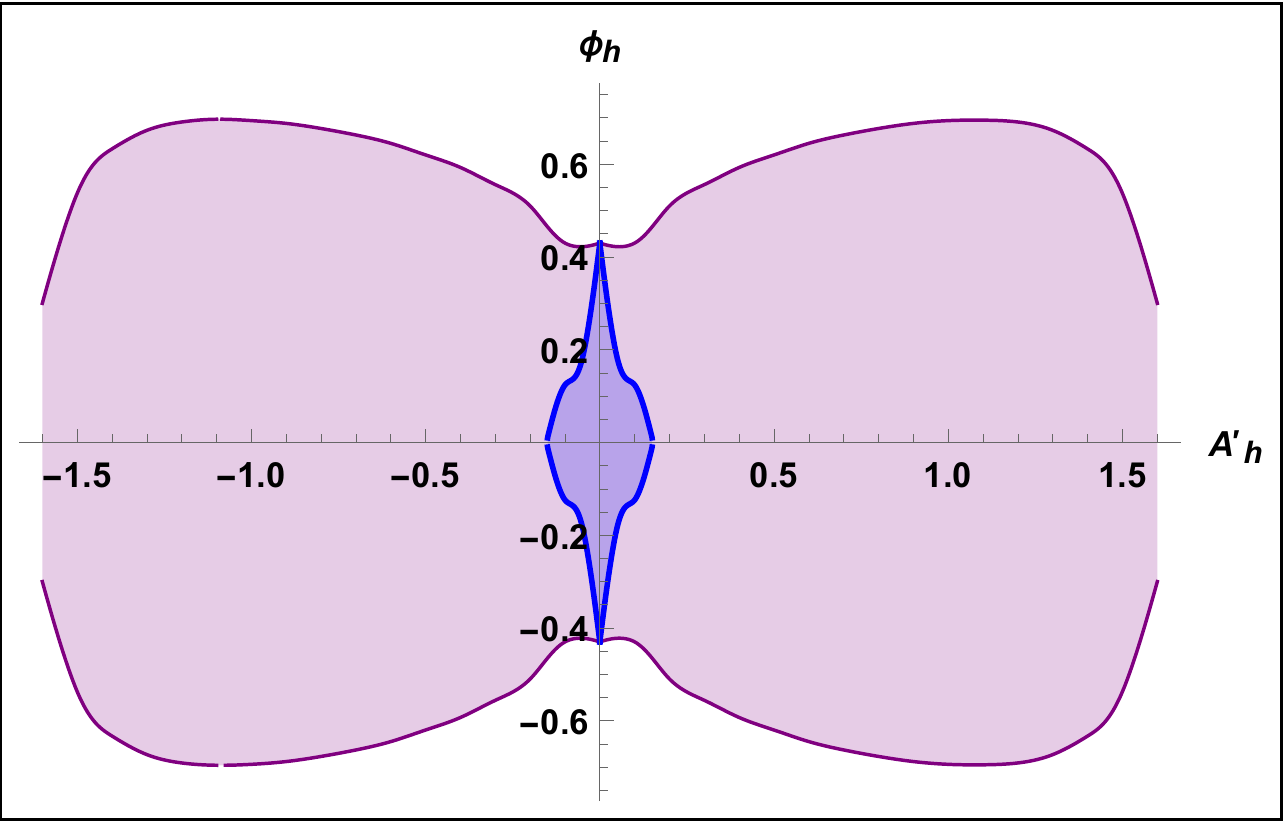}
\includegraphics[width=5.3cm,height=3cm]{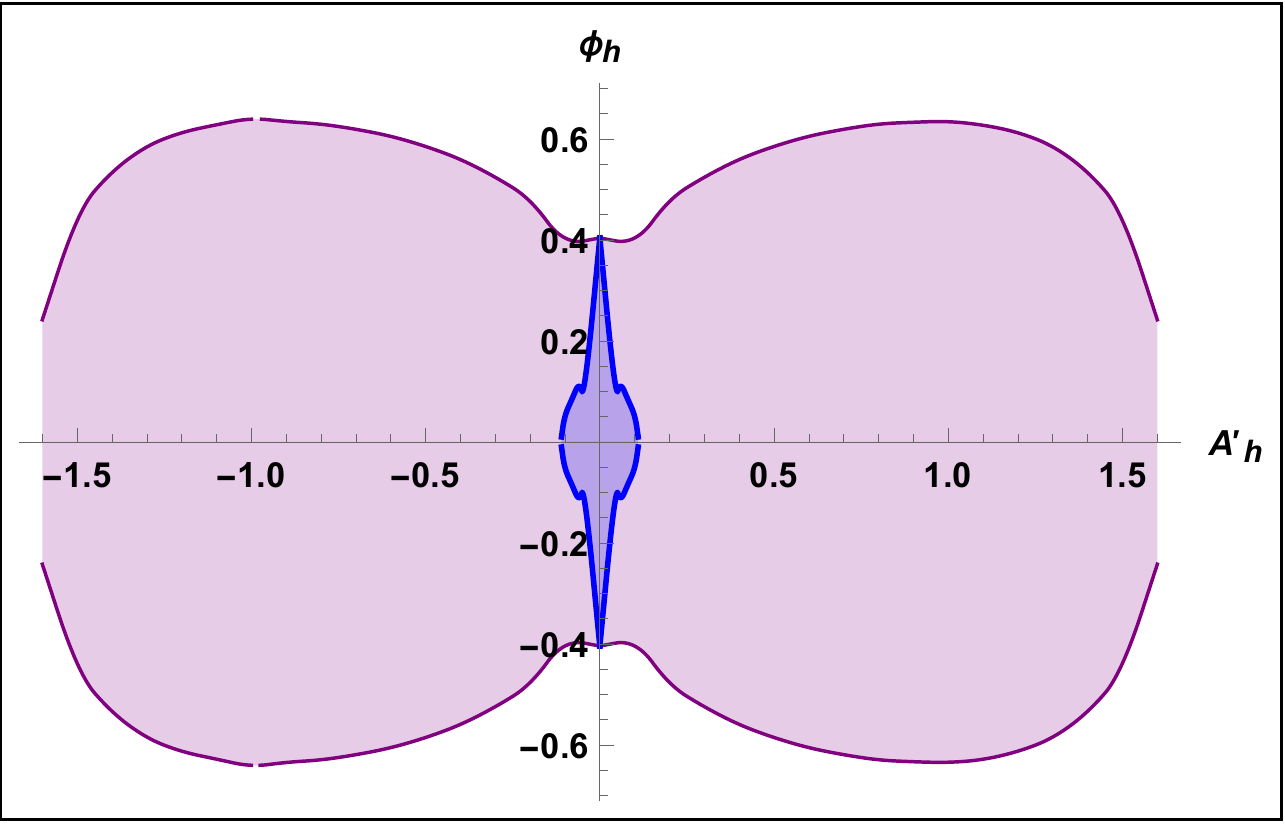}
 \caption{\footnotesize{Phase space ($A'_{h}$-$\phi_{h}$) of black hole solutions for different values of $q=0.4,0.6,0.8$ with $R_{0}=-0.03$ ($L=20$) (top row) and $R_{0}=-0.0048$ ($L=50$) (bottom row) for $F(R)=R-\frac{\mu^{4}}{R}$.}}\label{n}
 \end{figure}

\begin{figure}[!ht]
\includegraphics[width=8.2cm,height=4.5cm]{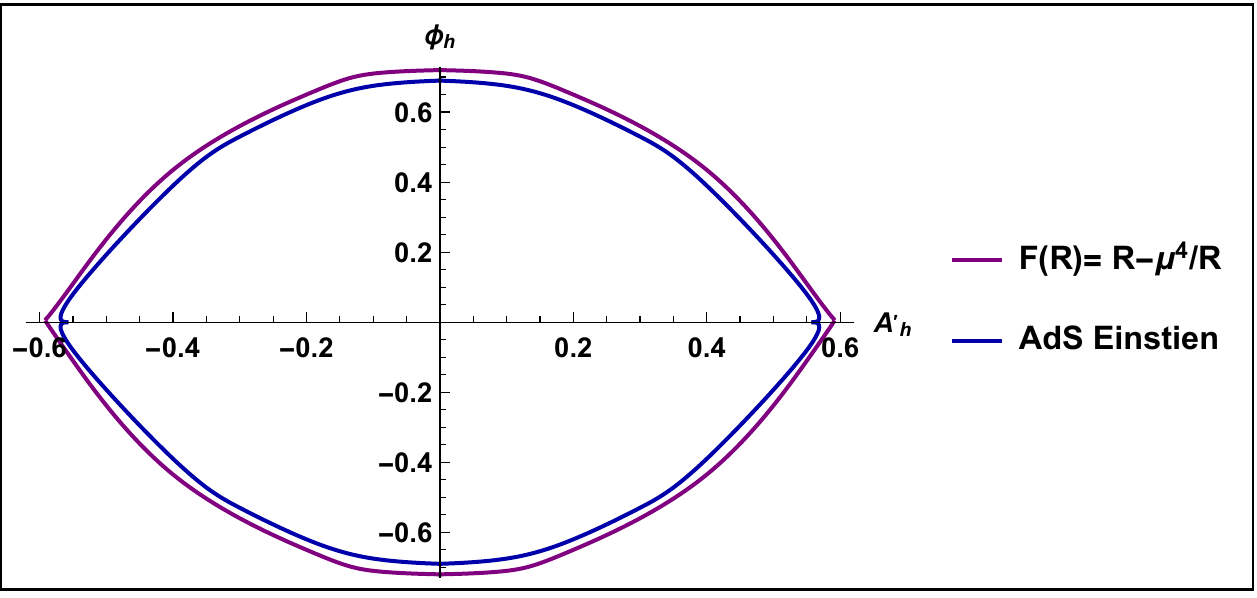}
\includegraphics[width=8.2cm,height=4.5cm]{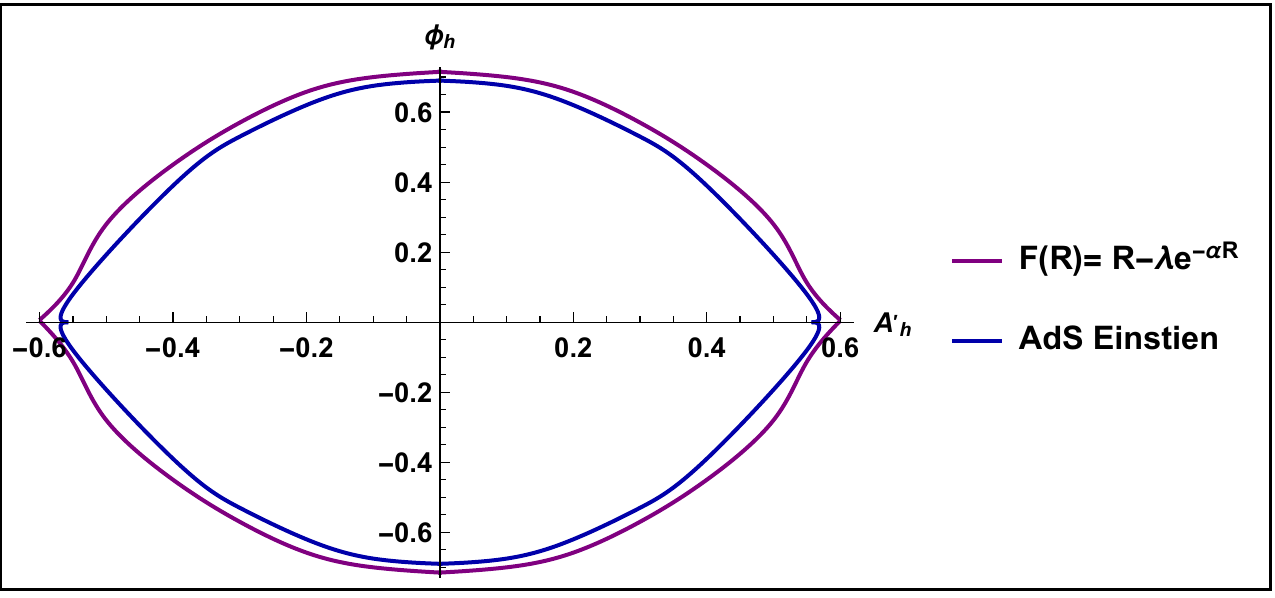}
\caption{\footnotesize{Comparison of phase space ($A'_{h}$-$\phi_{h}$) with $q=0.4$ and $R_{0}=-0.03$ ($L=20$) for black hole solutions with only one node in the framework of AdS Einstein-charged scalar field theory with $F(R)$ models. Left, $F(R)=R-\frac{\mu^{4}}{R}$ (left) and right, $F(R)=R-\lambda e^{-\alpha R}$ (right) with $\alpha=10$.} }\label{gr}
\end{figure}

So far, we have investigated the dependence of number of scalar field nodes on different parameters. Fig. \ref{fig2n} shows the solutions where the scalar field has only one node at the AdS boundary, $L$. Fig. \ref{fig20n} shows dependence of $q$ to the initial values of ($A'_{h}$ and $\phi_{h}$) for which the scalar field can have only one node at fixed AdS boundary. Such profiles are expected to be stable due to their low frequency and energy \cite{basou}. Given the importance of stable solutions, we expect to find at least one such solution around linear perturbations \cite{basou}.
As was mentioned before, one can confine the system by a natural or fictional reflecting boundary. In our cases, the time-like character of the AdS boundary mimics a natural reflecting boundary \cite{ads1, ads2, eli2}. Therefore, any wave scattered off the system will eventually reach spatial infinity (AdS boundary) and is reflected back again. Alternatively, one can use a fictional reflective wall as a new boundary located before the AdS radius and investigate the black hole stability at such a new boundary \cite{win, huang}.
As is well known, the phase space of a generic black hole is described by  $q$, $\phi_{h}$, $A'_{h}$ and $R_{0}$, whose possible hairy solutions can be obtained by fixing these parameters. To this end, the assumption $N'(r_{h})>0$ in equation (\ref{19}) leads to $|{A'_{h}}|<\sqrt{2(1+f_{R})-f_{R}R+f}$. As a consequence, $A'_{h}$ depends on the form of $F(R)$. Figs. \ref{fig200n} and \ref{n} show the parameter space of the static black hole solutions for different scalar field charges $q$. The solution space is indicated by the purple regions which show that the scalar field has at least one node up to the AdS boundary and the blue line shows solutions for which the scalar field has only one node at the AdS boundary with $N(r)>0$ for $r>r_{h}$. Note that there is no solution on the $A'_{h}=0$ axis since the black hole is uncharged. The blue regions are where the scalar field has no node, even at the AdS boundary and therefore are of no consequence. The solution space is a continuous region formed in the $\phi_{h}$-$A'_{h}$ plane for different values of the scalar field charges and shows that as $q$ or $L$ increase, the space of the static solution becomes smaller which is clearly seen for one node solutions. The reason is that when $q$ grows, the coupling between the scalar and the gauge field becomes stronger. As can be seen in Fig. \ref{gr} for the $F(R)$ models proposed here, the solution space with only one node at the AdS boundary is larger than that in the AdS Einstein-charged scalar field theory.

Summarizing, the $F(R)$ models presented here have static, spherically symmetric and asymptotically AdS solutions with scalar hair. The regularity of the solutions outside the event horizon, shown in  Fig. \ref{fig1}, confirms the existence of black hole solutions with a nontrivial scalar hair \cite{Heis}. It is clear that the scalar field profiles depend on the charge of the scalar field and AdS radius. We have considered the solutions of the hairy black hole family with one or more nodes up to the AdS boundary.
\section{Stability of solutions}
In the previous section, we numerically showed  that a fully coupled system of $F(R)$-Maxwell theory and a charged scalar field at the AdS boundary admits black hole solutions with scalar hair. In this configuration, the reflective character of AdS boundary provides the natural confining system. Since stable hairy solutions can be considered as a plausible final state of the superradiant instability \cite{win, gual, bosch, nami}, the next step is to verify the stability of the previous static solutions.

Following \cite{win}, to check the stability of the black hole under time evolution, we assume that the field variables, in addition to radial dependency, are time-dependent and are real except for the scalar field which was considered to be real in  the static case on account of the gauge freedom. Here however, it is considered to be complex. To simplify calculations, we define $\xi=N \sqrt{h}$. By manipulating equation (\ref{eq3}) and assuming that the field variable are functions of $(r,t)$ we obtain four dynamical equations

\begin{eqnarray}\label{22}
&&\frac{(1+f_{R})N'}{ N}+\frac{f_{R} R r-r f(R)}{2 N}-\frac{(1+f_{R})(1-N)}{N r}=-\frac{r}{2\xi^2}\left(|\dot{\phi}|^2+|\xi \phi'|^2+\right.\nonumber \\
&&\left.q^2|A_{0}|^2|\phi|^2+2qA_{0}\mbox{Im}(\phi\dot{\phi}^*)+NA_{0}'^2\right),
\end{eqnarray}

\begin{eqnarray}\label{23}
\frac{(1+f_{R})h'}{h}=\frac{r}{\xi^2}\left(|\dot{\phi}|^2+|\xi \phi'|^2+q^2 |A_{0}|^2|\phi|^2+2qA_{0}\mbox{Im}(\phi\dot{\phi}^*)\right),
\end{eqnarray}

\begin{equation}\label{24}
 \frac{(1+f_{R}) \xi'}{\xi}+\frac{f_{R}R r}{2 N}-\frac{r f(R)}{2 N} =-\frac{r N A_{0}'^2}{2\xi^2}+\frac{(1+f_{R})(1-N)}{Nr} ,
\end{equation}

\begin{equation}\label{25}
-\frac{(1+f_{R})\dot{N}}{N}=r \mbox{Re}(\dot{\phi}^*\phi')+r q A_{0}\mbox{Im}(\phi'^*\phi),
\end{equation}
\newline
where a dot denotes partial derivative with respect to $t$. Unlike the static case, there is an extra $(t r)$ component, equation (\ref{25}), resulting from  field equations (\ref{eq3}).
The Maxwell equations (\ref{eq4}) have two non-zero components ($t$ and $r$) whose corresponding dynamical equations become
\begin{equation}\label{26}
\frac{\xi}{r^2}\left(\frac{r^2A'_{0}}{h^{\frac{1}{2}}}\right)'=q^2|\phi|^2 A_{0}-q\mbox{Im}(\dot{\phi}\phi^*),
\end{equation}
\begin{equation}\label{27}
\frac{1}{r}\partial_{t}\left(\frac{{rA'_{0}}}{h^{\frac{1}{2}}}\right)=-q\mbox{Im}(\xi \phi'\phi^*).
\end{equation}
Let us now define a new variable  $\phi=\frac{\psi}{r}$  on account of the spherical symmetry of the scalar field. As for the Klein-Gordon equation (\ref{eq55}) we find
\begin{equation}\label{eq29}
-\ddot{\psi}+\left(\frac{\dot{\xi}}{\xi}+2iqA_{0}\right)\dot{\psi}+\xi(\xi \psi')'+\left(iq\dot{A_{0}}-\frac{\xi \xi'}{r}-iq\frac{\dot{\xi}}{\xi}A_{0}+q^2A_{0}^2\right)\psi=0.
\end{equation}
In what follows, we will focus attention on linear perturbations of the above equations and keep the terms to first order, for more detail see Appendix B.

\subsection{Linear perturbations}

To study the stability of the hairy black hole with an AdS boundary, we consider linear perturbations of dynamical equations around equilibrium solutions and define linear perturbations as $N(t,r)=\bar{N}(r)+\delta N(t,r)$ and similarly for $h(t,r)$, $\xi(t,r), \psi(t,r)$ and $A_{0}(t,r)$. Perturbation of $f(R(t,r))$ is defined by the expansion $f(R(t,r))=\bar{f}+\bar{f}_{R}\delta R(t,r)$ \cite{myon}, where $\bar{N}(r)$, $\bar{f} \equiv \bar{f}(R_{0})$ and $\bar{f_{R}}\equiv \bar{f}_{R}(R_{0})$ are equilibrium quantities and have radial dependency only, whereas $\delta N(t,r)$ and  $\delta R(t,r)$ show perturbation parts. Substitution of the above definitions in (\ref{22}-\ref{eq29}) will now result in equations for the perturbed quantities which we will discuss below.

The perturbed scalar field  $\delta\psi$ is a complex quantity but other perturbed fields are real. We decompose $\delta \psi$ into real and imaginary parts as $\delta \psi=\delta u+i\delta \dot{w}$ where $\delta u$ and $\delta w$ are real. Note that this decomposition was chosen to satisfy the dynamical perturbation equations (see equation (\ref{43}) in Appendix B).
One gets three coupled perturbed equations in terms of $\delta A$, $\delta u$ and $\delta w$. The first two are dynamical involving time derivatives as follows, see Appendix B for more details.
\begin{eqnarray}\label{32}
&&\delta \ddot{u}-\bar{\xi}^2 \delta u''-\bar{\xi} \bar{\xi}'\delta u'+\left[3 q^2 \bar{A_{0}}^2+\frac{\bar{\xi} \bar{\xi}'}{r}+\frac{ \bar{N}\left(\frac{\bar{\psi}}{r}\right)'^2 }{1+\bar{f}_{R}}\left(\frac{r^2 \bar{A_{0}}'^2}{2 (1+\bar{f}_{R})}-\frac{r^2 \bar{h} \bar{f}}{2( 1+\bar{f}_{R})}+\frac{r^2 \bar{h} \bar{f}_{R} \bar{R_{0}}}{2( 1+\bar{f}_{R})}-\bar{h}\right)\right]\delta u\nonumber \\
&&+ 2 q \bar{A_{0}} \bar{\xi}^2 \delta w''+q \bar{N}\bar{A_{0}}\left[2 \sqrt{\bar{h}}\bar{\xi}'+\frac{\bar{\psi} \left(\frac{\bar{\psi}}{r}\right)'}{1+\bar{f}_{R}}\left(-\frac{\bar{N}\bar{h} \bar{A_{0}}' }{\bar{A_{0}}}-\frac{r \bar{A_{0}}'^2}{2(1+\bar{f}_{R})}-\frac{\bar{h}\bar{f}_{R} \bar{R_{0}} r}{2 (1+\bar{f}_{R})}+\frac{\bar{h}}{r}+\frac{\bar{h}\bar{f} r}{2 (1+\bar{f}_{R})}\right)\right] \nonumber \\
&&\times\delta w' +q \bar{A_{0}}\left[ 2 q^2 \bar{A_{0}}^2-\frac{2 \bar{\xi} \bar{\xi}'}{r}+\frac{\bar{\xi} \bar{\psi}' \left(\frac{\bar{\psi}}{r}\right)'}{1+\bar{f}_{R}}\left(\frac{\bar{\xi} \bar{A_{0}}'}{\bar{A_{0}}}-\bar{\xi}'-\frac{\bar{\xi}}{r }\right)\right]\delta w=0,
\end{eqnarray}
\begin{eqnarray}\label{33}
&&\delta \ddot{w}-\bar{\xi}^2\delta w''+\left[\frac{q^2 \bar{A_{0}} \bar{\psi}^2}{r^2  \bar{A_{0}}'}\left(\frac{r \bar{A_{0}}' \bar{A_{0}}}{1+\bar{f}_{R}}+ \bar{N}\bar{h}\right)-\bar{\xi} \bar{\xi}'\right] \delta w'-\left[\frac{q^2 \bar{A_{0}} \bar{\psi} \bar{\psi}'}{r^2  \bar{A_{0}}'} \left(\frac{ r \bar{A_{0}} \bar{A_{0}}'}{1+\bar{f}_{R}}+\bar{N}\bar{h}\right)+q^2 \bar{A_{0}}^2\right.\nonumber \\
&&\left.-\frac{\bar{\xi} \bar{\xi}'}{r}\right] \times \delta w -q \bar{A_{0}}\left(2+\frac{\bar{\psi} \left(\frac{\bar{\psi}}{r}\right)'}{1+\bar{f}_{R}}\right)\delta u-q \bar{\psi} \delta A+\frac{q \bar{A_{0}} \bar{\psi}}{\bar{A_{0}}'} \delta A'=0,
\end{eqnarray}
while the third equation is a constraint
\begin{eqnarray}\label{34}
&&\frac{q \bar{\psi}}{r}\left(\frac{\bar{A_{0}}}{1+\bar{f}_{R}}+\frac{\bar{N}\bar{h}}{r \bar{A_{0}}'}\right)\delta w''+ q \bar{A_{0}} \bar{\psi} \left[-\frac{q^2 \bar{h} \bar{\psi}^2}{r^4 \bar{A_{0}}'^2}+\frac{\bar{\xi}'}{(1+\bar{f}_{R}) \bar{\xi} r}+\frac{\sqrt{\bar{h}}\bar{\xi'}}{r^2 \bar{A_{0}} \bar{A_{0}}'}\right]\delta w'+\frac{q  \bar{A_{0}} \bar{\psi}}{r^2}\left[\frac{r q^2 \bar{A_{0}}^2}{(1+\bar{f}_{R}) \bar{\xi}^2}\right.\nonumber\\
&&\left.+\frac{q^2 \bar{A_{0}}}{\bar{N} \bar{A_{0}}'}-\frac{\bar{\xi}'}{(1+\bar{f}_{R}) \bar{\xi}}-\frac{\sqrt{\bar{h}}\bar{\xi}'}{r \bar{A_{0}} \bar{A_{0}}' }+\frac{q^2 \bar{h} \bar{\psi}\bar{\psi}'}{r^2 \bar{A_{0}}'^2}\right]\delta w-\frac{\left(\frac{\bar{\psi}}{r}\right)'}{1+\bar{f}_{R}}\delta u'-\left[ \frac{\left(\frac{\bar{\psi}}{r}\right)'}{1+\bar{f}_{R}}\left(\frac{1}{r}+\frac{\bar{\xi}'}{\bar{\xi}}\right) + \frac{\left(\frac{\bar{\psi}}{r}\right)''}{1+\bar{f}_{R}}\right]\delta u\nonumber\\
&&+\frac{\delta A''}{\bar{A_{0}}'} -\frac{\bar{A_{0}}''}{\bar{A_{0}}'^2}\delta A'=0,
\end{eqnarray}
where $\bar{f}_{RR}$ denotes $\frac{d\bar{f}_{R}}{dR}$.
\subsection{Boundary conditions}
Perturbation modes must satisfy boundary conditions at the event horizon and reflective boundary (AdS boundary). They are assumed to be of the time-periodic form for solving perturbations equations and  need to satisfy ingoing wave-like conditions at the event horizon
\begin{eqnarray}\label{35}
&&\delta u(t,r)=\mbox{Re}[ e^{-i\omega (t+r_{*})} U(r)],\nonumber \\
&&\delta w(t,r)=\mbox{Re}[ e^{-i\omega (t+r_{*})} W(r)],\nonumber \\
&&\delta A(t,r)=\mbox{Re}[ e^{-i\omega (t+r_{*})} A(r)],
\end{eqnarray}
where $U$, $W$ and $A$ are complex functions that depend on the radial coordinate and have regular Taylor expansions near the horizon
\begin{eqnarray}\label{305}
&&U(r)=U_{0}+U_{1}(r-r_{h})+U_{2} (r-r_{h})^2/2+...\nonumber \\
&&W(r)=W_{0}+W_{1}(r-r_{h})+W_{2} (r-r_{h})^2/2+...,\nonumber \\
&&A(r)=A_{0}+A_{1}(r-r_{h})+A_{2} (r-r_{h})^2/2+....
\end{eqnarray}
By substituting expansion series (\ref{305}) and ingoing wave conditions at the event horizon in  perturbation equations (\ref{32}-\ref{34}), we obtain $U_{1}$, $W_{1}$ and $A_{1}$ in terms of $U_{0}$, $W_{0}$ and $\omega$.
\begin{eqnarray}\label{306}
&&U_{1}=\frac{\frac{-\alpha_{2}}{r_{h}}U_{0}+2q\omega^2\alpha_{1}W_{0}}{2i\omega-\alpha_{2}},\nonumber \\
&&A_{1}=\frac{-q\phi_{h}\omega^2A'_{h}(\frac{\alpha_{1}}{1+f_{R}}+\frac{1}{r_{h}A'_{h}})W_{0}}{i\omega+\frac{\omega^2}{\alpha_{2}}},\nonumber \\
&&W_{1}=\frac{-2q\alpha_{1}U_{0}+\left(\frac{\alpha_{2}}{r_{h}}-\frac{i\omega q^2\phi_{h}^2\alpha_{1}}{A'_{h}}-\frac{i\omega q^2r_{h}\phi_{h}^2 {\alpha_{1}}^2}{1+f_{R}}\right)W_{0}-\frac{iqr_{h}\phi_{h}\alpha_{1}\omega}{A'_{h}\alpha_{2}}A_{1}}{\alpha_{2}-2i\omega},
\end{eqnarray}
where
\begin{eqnarray}\label{307}
&&\alpha_{1}=\frac{2r_{h}A'_{h}(1+f_{R})}{-{r_{h}}^2\left({A'_{h}}^2-f+f_{R}R_{0}\right)+2(1+f_{R})}\nonumber \\
&&\alpha_{2}=-\frac{r_{h}}{2(1+f_{R})}\left({A'_{h}}^2-f+f_{R}R_{0}\right)+\frac{1}{r_{h}},
\end{eqnarray}
To avoid having a singularity in perturbation equations at the event horizon, we must set $A_{0}=0$.
It is worth noting that at the reflective boundary, we need both real and imaginary parts of $U$ and $W$ to disappear. We set $W_{0}=1$ and fix the scale of the perturbation since perturbation equations and boundary conditions are linear \cite{win}. Then only $U_{0}$ and $\omega$ are free parameters and perturbation equations with boundary conditions at the event horizon define an eigenvalue problem with eigenvalue $\omega$ which we seek to find. In the case of $\mbox{Im}(\omega)\leq 0$, the black hole solutions are linearly stable and for $\mbox{Im}(\omega)>0$, the black hole solutions are unstable.
\subsection{Numerical solutions and results}
In subsection \ref{sub1}, the black hole solution space was described with parameters $q$, $\phi_{h}$, $A'_{h}$ and $R_{0}$. By fixing these quantities we solve equations (\ref{18}-\ref{21}) numerically, deriving the static black hole solutions. Fig. \ref{fig2} shows that for the small $q$ or $L$, the number of nodes is reduced up to the reflective boundary and Fig. \ref{fig2n} shows the family of black hole solutions that have only one node at the AdS boundary. For a careful examination of the stability of the black hole for both families of solutions (the static solutions of scalar field have one node or more up to the AdS boundary), we integrate perturbation equations (\ref{32}-\ref{34}) using boundary conditions as initial conditions. We seek parameters $U_{0}$ and $\omega$ so as to satisfy the condition of the perturbed scalar field at the reflective boundary (the real and imaginary parts of the perturbed scalar field should vanish at the reflective boundary). So $U_{0}$ and $\omega$ play the role of shooting parameters in the shooting method.

\begin{figure}[!ht]
\includegraphics[width=5.3cm,height=3cm]{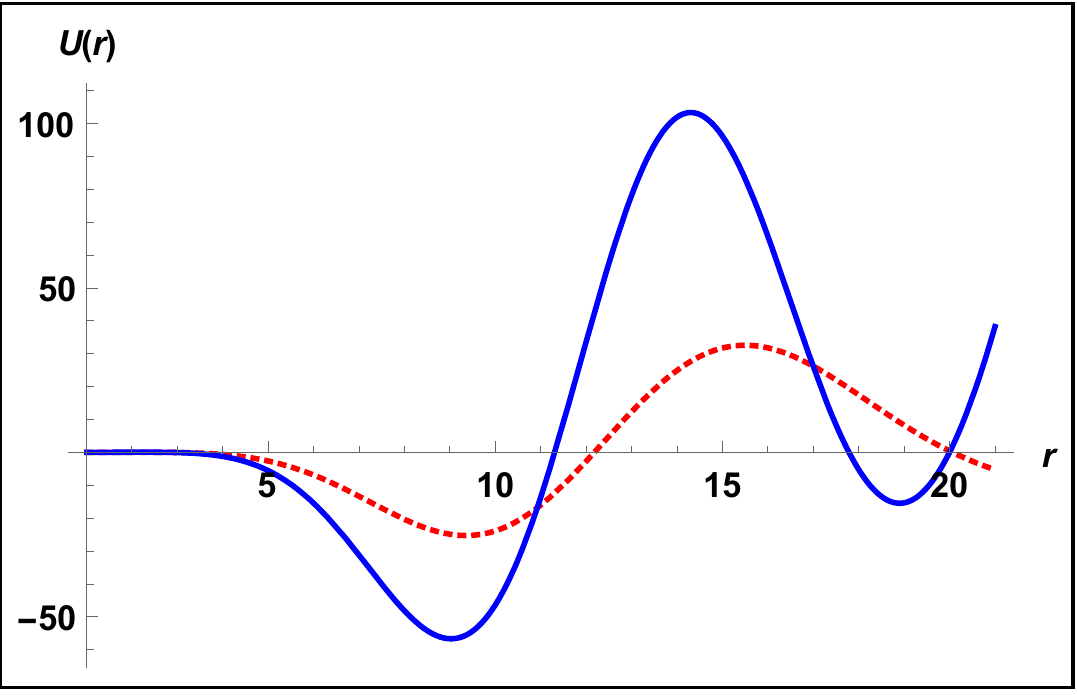}
\includegraphics[width=5.3cm,height=3cm]{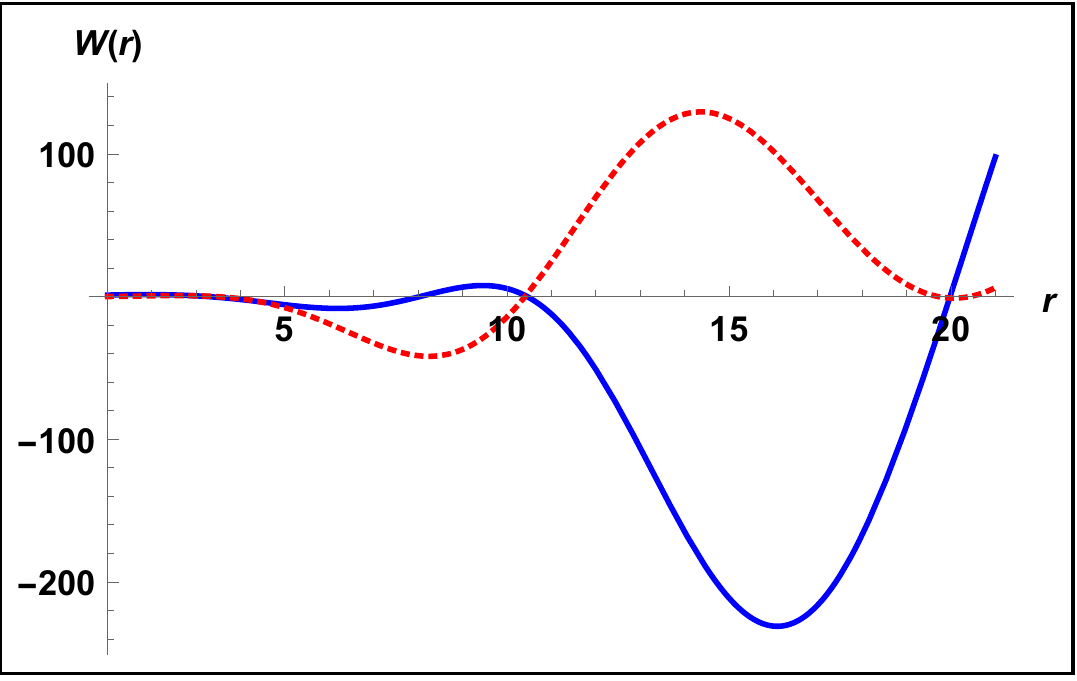}
\includegraphics[width=5.3cm,height=3cm]{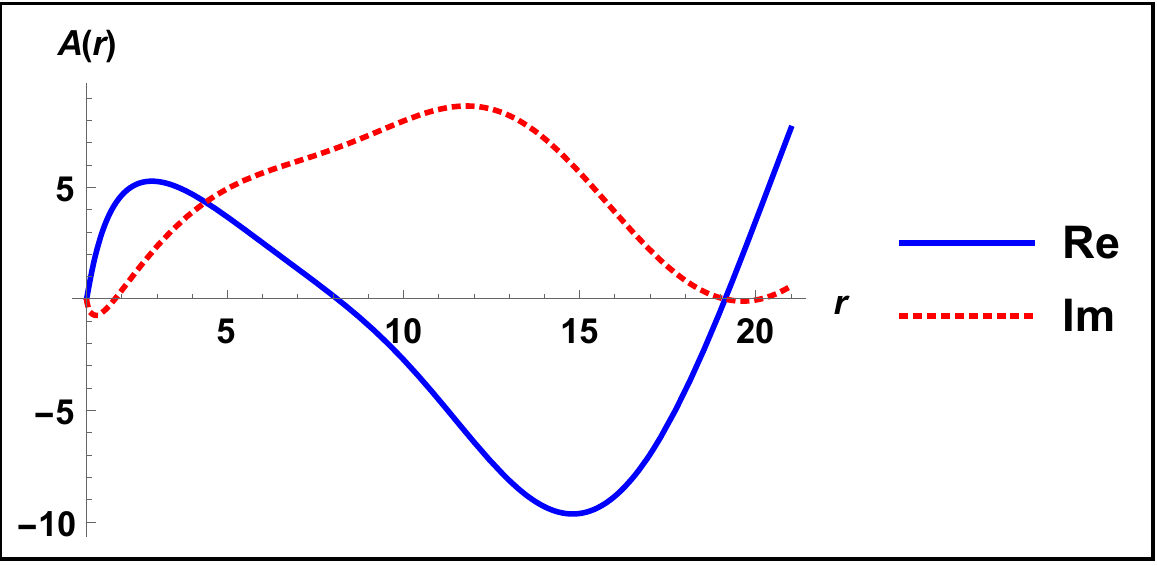}
 \caption{\footnotesize{Behavior of the three perturbation functions $U(r)$, $W(r)$ and $A(r)$ for $F(R)=R-\lambda e^{-\alpha R}$ with q =0.4, $A_{h}'=0.47$, $\phi_{h}=0.35$, $U_{0}=0.0015+0.001i$, $R_{0}=-0.03$ ($L=20$), $\alpha=10$ and $\omega=0.194 - 0.00097i $. }}\label{fig7}
\end{figure}

The three perturbation functions $U(r)$, $W(r)$ and $A(r)$ are shown in Fig. \ref{fig7} for $F(R)=R-\lambda e^{-\alpha R}$. The values $A'_{h}$ and $\phi_{h}$ are selected according to the phase space, Figs. \ref{fig200n} and \ref{gr}, such that we can find static solutions of the scalar field with just one node at the AdS boundary. Also, the shooting parameters, $U_{0}=0.0015+0.001i$ and $\omega=0.194 - 0.00097i $ are so identified  as to satisfy the vanishing of $U$ and $W$ modes at the reflective boundary. These perturbation modes decay exponentially since frequencies satisfy $\mbox{Im}(\omega)<0$. Such black hole solutions are then stable at the linear level. For $F(R)=R-\frac{\mu^{4}}{R}$ with $q=0.3$, $R_{0}=-0.03$, $A'_{h}=0.65$ and $\phi_{h}=0.4$, we find a configuration of the scalar field with just one node at the AdS boundary and shooting parameters  $U_{0}=0.026+0.0155i$ and $\omega=0.193 - 0.001158i$, leading to the result and figure similar to that for $F(R)=R-\lambda e^{-\alpha R}$ .
\begin{figure}[!ht]
\includegraphics[width=8.2cm,height=4.5cm]{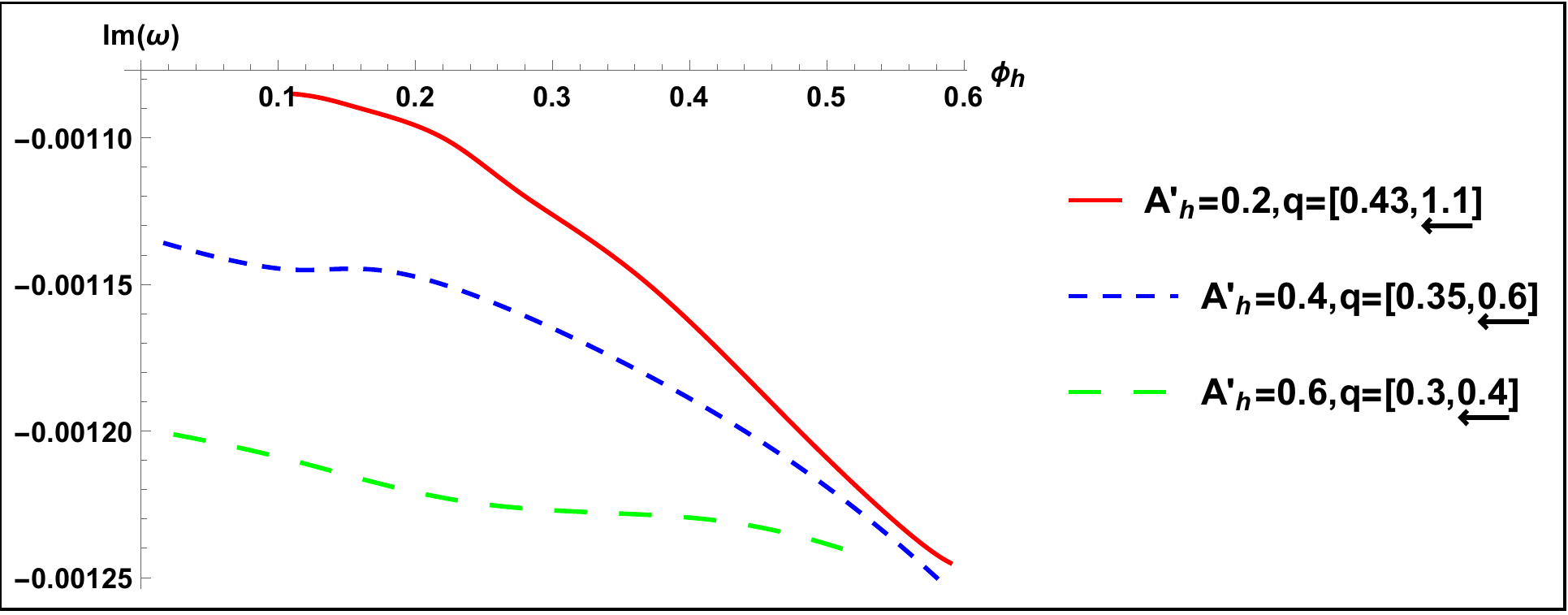}
\includegraphics[width=8.2cm,height=4.5cm]{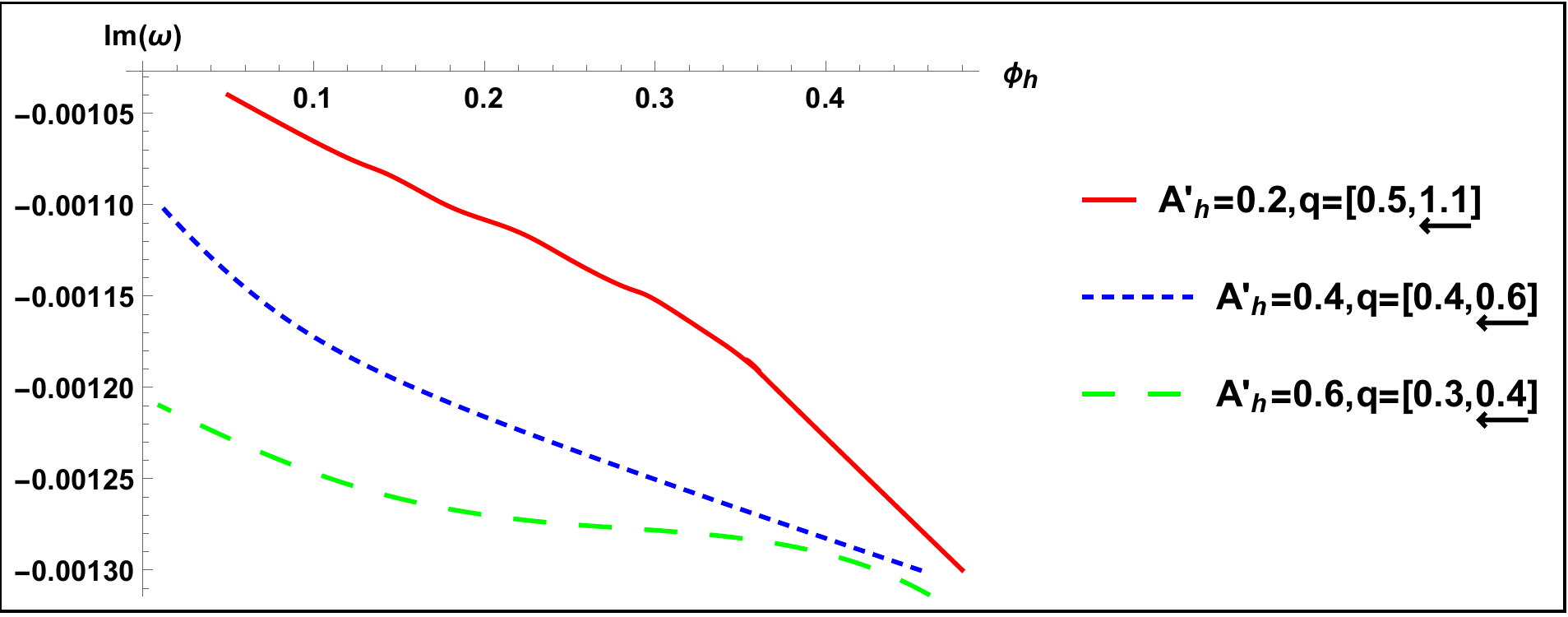}\\
\includegraphics[width=8.2cm,height=4.5cm]{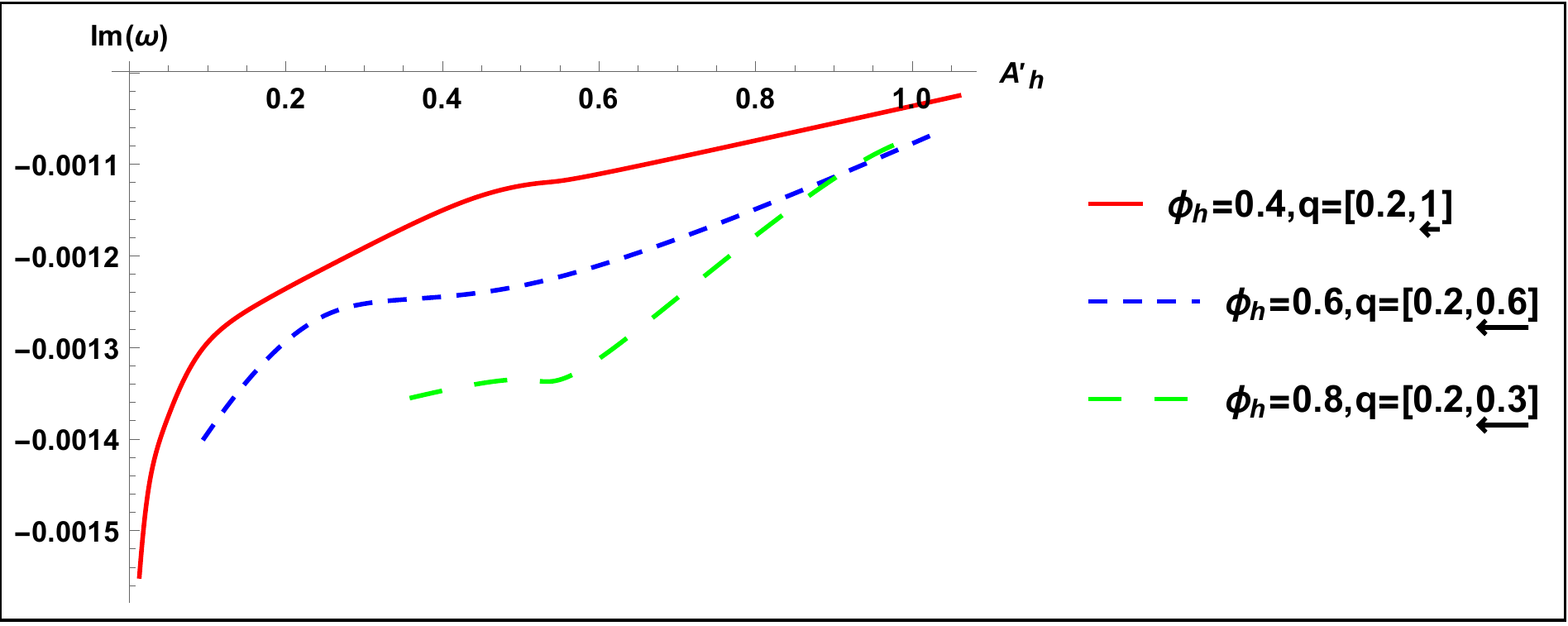}
\includegraphics[width=8.2cm,height=4.5cm]{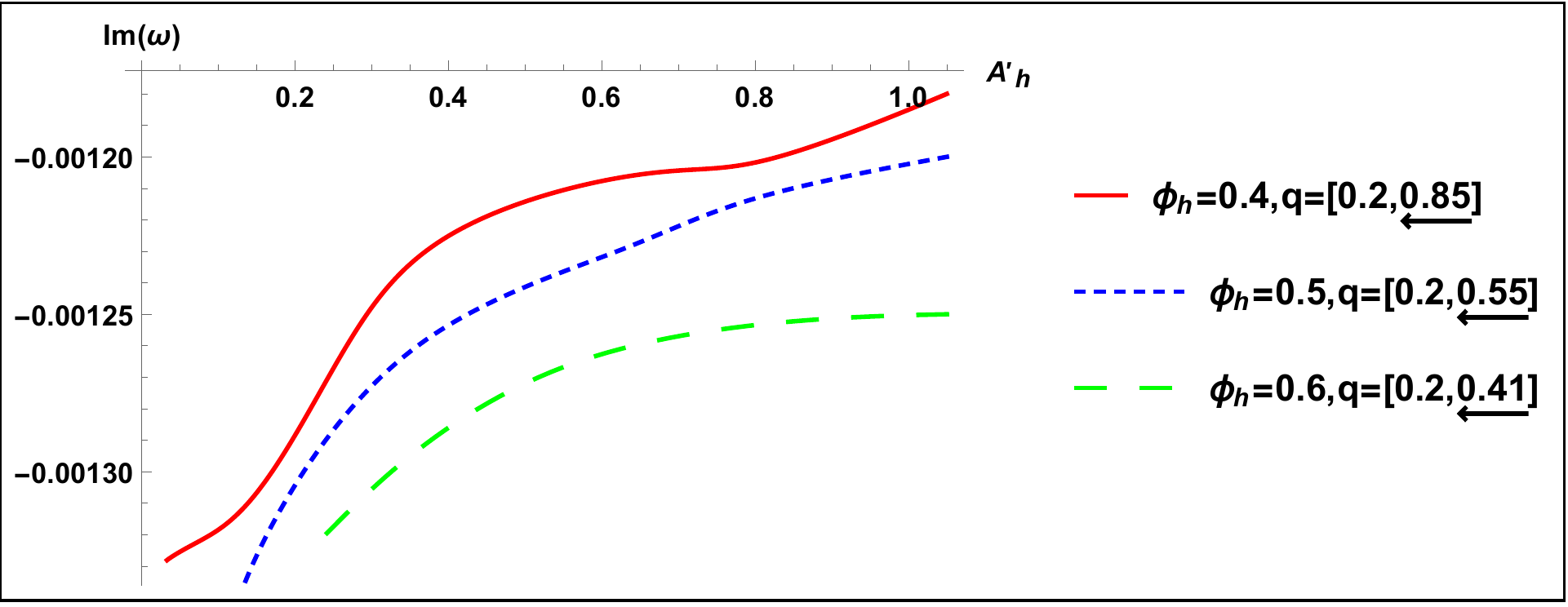}
\caption{\footnotesize{ The imaginary parts of perturbation modes plotted as a function of  $\phi_{h}$, top row, with different values for $q$ and $A'_{h}$, and bottom row, $A'_{h}$ with different values for $q$ and $\phi_{h}$ when the static solutions of the scalar field have only one node at the AdS boundary for, Left: $F(R)=R-\frac{\mu^{4}}{R}$ and, Right: $F(R)=R-\lambda e^{-\alpha R}$ with $\alpha=10$. We fix the value of $R_{0}=-0.03$ ($L = 20$) for the AdS radius. Here, the perturbation modes decay exponentially in time since the imaginary frequency is negative.}}\label{fig2r}
\end{figure}

In Fig. \ref{fig2r}, the imaginary parts of perturbation modes, when static solutions of the scalar field have only one node at the AdS boundary, are shown as functions of $A'_{h}$ and $\phi_{h}$. In Fig. \ref{fig2r}, by varying $q$ and $\phi_{h}$ for fixed values of $R_{0}$ and $A'_{h}$ or  $q$ and $A'_{h}$ for fixed values of $R_{0}$ and $\phi_{h}$, one can find static solutions of the scalar field with only one node at AdS boundary, Fig. \ref{fig20n}. Since for all values of $A'_{h}$ and $\phi_{h}$ with one node phase space of solutions, the $\mbox{Im}(\omega)$ are negative \footnote{The hairy black hole for which the scalar field has only one node ($n=0$) at the AdS boundary is described as a weakly interacting mix of a  RNAdS black hole and a condensate of the ground state of the
scalar field \cite{basou}.}, we conclude that perturbation modes decay exponentially in time, leading to stable hairy black holes that can be signified as the possible endpoint of superradiant instability for the models of $F(R)$ discussed here (as perturbation modes fail to become superradiant, the final state is completely specified by a stable hairy black hole). As can be seen in this figure, $\mbox{Im}(\omega)$ decrease by increasing $\phi_{h}$ for fixed $A'_{h}$ .

\begin{figure}[!ht]
\includegraphics[width=8.2cm,height=4.5cm]{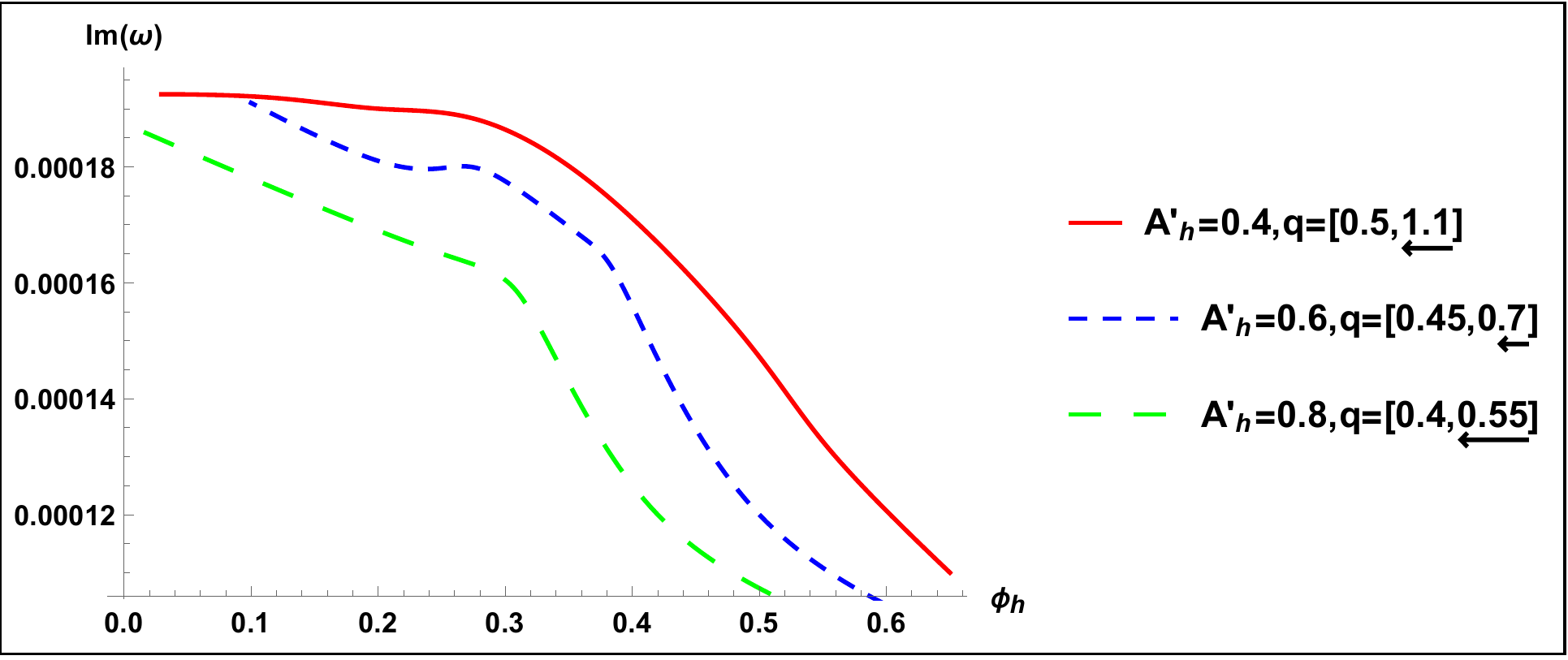}
\includegraphics[width=8.2cm,height=4.5cm]{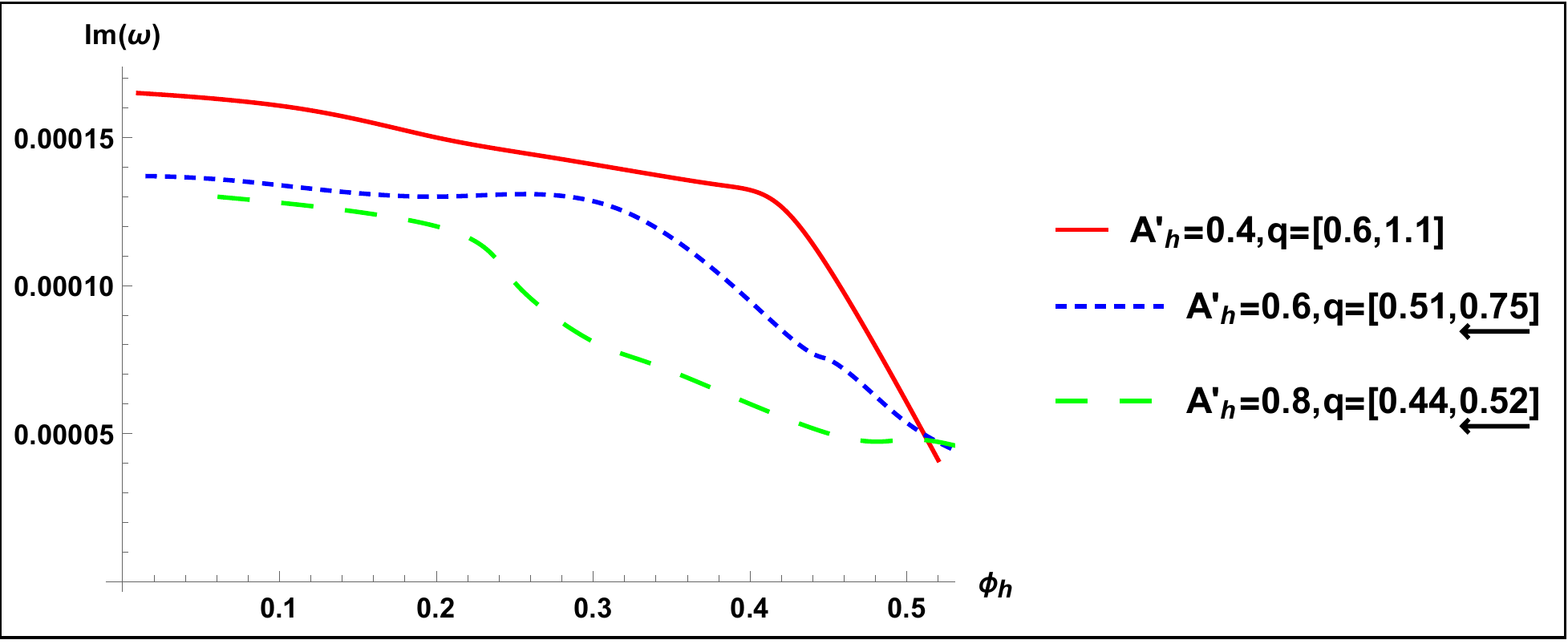}
 \caption{\footnotesize{The imaginary parts of perturbation modes when the static solutions of the scalar field have two nodes
up to the AdS boundary, plotted against $\phi_{h}$ with fixed AdS radius $R_{0}=-0.03$ (L = 20) and different values of $q$ and $A'_{h}$. Here, the
perturbation modes grow exponentially in time since the imaginary frequency is positive. Left: $F(R)=R-\frac{\mu^{4}}{R}$. Right: $F(R)=R-\lambda e^{-\alpha R}$ with $\alpha=10$. }}\label{fig20r}
\end{figure}

Fig. \ref{fig20r} illustrates plots of $\mbox{Im}(\omega)$ for both models as a function of $\phi_{h}$ for different values of $A'_{h}$ where static solutions of the scalar field have two nodes up to the AdS boundary. To find such solutions, $q$ and $\phi_{h}$ are varied for fixed values of $A'_{h}$ and $R_{0}$. As can be seen in Fig. \ref{fig20r}, the $\mbox{Im}(\omega)$ is positive \footnote{The solutions
with more nodes correspond to weakly interacting mix of the RNAdS
black hole and an excited state of the scalar field. They
are unstable and presumably decay to the ground state of the hairy black
hole \cite{basou}.} which indicates exponentially growing modes in time and thus hairy black holes are unstable for more than one node up to the AdS boundary.
\section{Conclusions}
When a charged scalar field impinges on a $F(R)$-Maxwell black hole and scatters off it in an asymptotically AdS space-time, the reflective AdS boundary behaves as a natural confining system that leads to superradiant instability.

In this work, we have found static solutions in such a setting with a scalar hair in $F(R)$ theory in a numerical fashion which shows that the scalar field fluctuates around zero. We  also considered profiles of the scalar field that have only one node at the AdS boundary. We derived the phase space of black hole solutions and showed that the static solutions of the scalar field have only one node at the AdS boundary due to the selected models of $F(R)$ whose phase space is larger than that in the AdS Einstein-charged scalar field theory. We also investigated the stability of hairy black holes in this context. To this end, we considered dynamical equations to first order in perturbations, the result of which was  the three coupled equations (\ref{32}-\ref{34}) which could be integrated numerically using the shooting method for which, in effect, the boundary conditions at the horizon can be considered as initial conditions. We also derived values of the frequency in such a way as to satisfy the vanishing of the perturbed scalar modes at the reflective boundary, that is, such a frequency can be considered as a shooting parameter.

We showed that if the scalar field has more than one radial node up to the AdS boundary, the sign of the imaginary part of the frequency becomes positive which leads to instability of the system. If however, the scalar field has only one node at the AdS boundary, the sign of the imaginary part of the frequency becomes negative for which the perturbation modes decay exponentially in time and the hairy black hole becomes stable so that perturbation modes fail to become superradiant. This means that under such conditions stable hairy black holes can be considered as a possible endpoint of superradiant instability. Our results are consistent with the results presented in \cite{win} where a fictional mirror, instead of AdS boundary, is used in a RN black hole and a charged scalar field set up. Finally, two points are worth mentioning regarding the analysis presented in this paper: first, the phase space enhancement of the solutions with respect to general relativity is marginal and second, the stability analysis is restricted to spherically symmetric perturbations and all conclusions refer to this case.

In summary, we have investigated superradiant instability of charge black holes for selected models of $F(R)$. The main motivation was to select models of $F(R)$ in such a way as to create specific geometry in the form of an asymptotically AdS space-time which would lead to superradiant instability, often referred to as a black hole bomb.
\section*{Acknowledgements}
M. Honardoost would like to thank Iran National Science Foundation (INSF) and Research Council of Shahid Beheshti University for financial support. We also thank M. Khodadi for a careful reading of the manuscript and helpful comments.
 \vspace{3mm}
 \appendix
 \section{ Appendix A}
 By assuming constant curvature, the field equation (\ref{eq3}) in vacuum becomes
 \begin{equation}\label{a1}
(1+f_{R})R_{\mu\nu}-\frac{1}{2}(R+f(R))g_{\mu\nu}=0.
\end{equation}
The trace of (\ref{a1}) leads to
\begin{equation}\label{a2}
(1+f_{R})R-2(R+f(R))=0.
\end{equation}
Now, substituting for $f(R)$ in (\ref{a2}), one gets $R_{0}=\pm \sqrt{3} \mu^2 $ for $f(R)=-\frac{\mu^{4}}{R}$ and $\lambda= \frac{R_{0} e^{\alpha R_{0}}}{2+\alpha R_{0}}$ for $f(R)=-\lambda e^{-\alpha R}$, where $\alpha$ is a free parameter, representing solutions which correspond to a topological Schwarzschild-AdS(dS) black hole. In the numerical solution, we fix $R_{0}$ to get $\mu^{2}$ and $\lambda$. In \cite{cruze}, by considering thermodynamics of black holes in an AdS space-time, the authors found that the necessary conditions for $F(R)$ to support AdS black holes are
\begin{equation}\label{3a}
R+f(R)<0,
\end{equation}
\begin{equation}\label{3aa}
1+f_{R}>0.
\end{equation}
In addition, the condition $\frac{d^{2}f}{dR^{2}}$ guarantees that there is no tachyonic instability. These conditions may impose extra restrictions on $R_{0}$. In the first model, only $R_{0}=\mp \sqrt{3} \mu^2$ would allow compliance with a topological Schwarzschild-AdS black hole. However, in the second model with the assumption $\alpha >0$, these conditions require $-\frac{1}{\alpha}<R_{0}<0$.

\section{ Appendix B}
Perturbations of dynamical equations (\ref{22}-\ref{eq29}), to first order, can be derived using definition of the perturbed quantities  defined in the text with the result
\begin{eqnarray}\label{38}
&&\frac{(1+\bar{f}_{R})\delta N'}{\bar{N}}-\left[\frac{1+\bar{f}_{R}}{\bar{N}}\left(\bar{N'}-\frac{1}{r}\right)-\frac{r \bar{A_{0}'}^2}{2 \bar{N}\bar{h}}-\frac{\bar{f}r}{2\bar{N}}+\frac{\bar{f}_{R}r\bar{R}}{2\bar{N}}\right]\frac{\delta N}{\bar{N}}=\left(\frac{q^2 \bar{A_{0}}^2 \bar{\psi}^2}{r \bar{\xi}^2}+\frac{r \bar{A_{0}}'^2}{\bar{N}\bar{h}}\right)\frac{\delta \xi}{\bar{\xi}}\nonumber \\
&&-\frac{q^2 \bar{A_{0}} \bar{\psi}^2}{r \bar{\xi}^2}\delta A-\frac{r \bar{A_{0}}'}{\bar{N}\bar{h}} \delta A'-\bar{f}_{RR} \left(\frac{1}{\bar{N}r}(\bar{N}-1)+\frac{ r \bar{R_{0}}}{2 \bar{N}}+\frac{\bar{N}'}{\bar{N}}\right) \delta R-\frac{i q \bar{A_{0}} \bar{\psi}}{2 r \bar{\xi}^2}(\delta \dot{\psi}-\delta \dot{\psi}^\ast) -\left(\frac{\bar{\psi}}{2 r}\right)' \nonumber \\
&&\times(\delta \psi'+\delta \psi'^\ast)+\left[\frac{\left(\frac{\bar{\psi}}{r}\right)'}{2 r}-\frac{q^2\bar{A_{0}}^2 \bar{\psi}}{2 r \bar{\xi}^2}\right](\delta \psi+\delta \psi^\ast),
\end{eqnarray}
\begin{eqnarray}\label{39}
&&\frac{(1+\bar{f}_{R})\delta h'}{\bar{h}}-\frac{(1+\bar{f}_{R})\bar{h'}\delta h}{\bar{h}^2}+\frac{\bar{f}_{RR}\bar{h'}\delta R}{\bar{h}}=\frac{i q \bar{A_{0}} \bar{\psi}}{r \bar{\xi}^2}(\delta \dot{\psi}-\delta \dot{\psi}^\ast) -\left[\frac{1}{r}\left(\frac{\bar{\psi}}{r}\right)'-\frac{q^2 \bar{A_{0}}^2 \bar{\psi}}{r \bar{\xi}^2}\right](\delta \psi+\delta \psi^\ast) \nonumber \\
&&+\left(\frac{\bar{\psi}}{r}\right)'(\delta \psi'+\delta \psi'^\ast)+\frac{2 q^2 \bar{A_{0}} \bar{\psi}^2}{r \bar{\xi}^2} \delta A-\frac{2 q^2 \bar{A_{0}}^2 \bar{\psi}^2}{r \bar{\xi}^3}\delta \xi,
\end{eqnarray}
\begin{eqnarray}\label{40}
&&\frac{(1+\bar{f}_{R})\delta \xi'}{\bar{\xi}}=-\left[\frac{r\bar{A_{0}'}^2}{2 \bar{h}}+\frac{1}{r}(1+\bar{f_{R}})+\frac{\bar{f} r}{2 }-\frac{\bar{f}_{R} \bar{R_{0}}r}{2}\right]\frac{\delta N}{\bar{N}^2}+\left[\frac{(1+\bar{f}_{R})\bar{\xi}'}{\bar{\xi}}+\frac{r\bar{A_{0}'}^2}{\bar{N} \bar{h}}\right]\frac{\delta \xi}{\bar{\xi}}-\left(\frac{1}{r}\right.\nonumber \\
&&\left.-\frac{1}{\bar{N}r}+\frac{r\bar{R}}{2 \bar{N}}+\frac{\xi'}{\xi}\right)\bar{f}_{RR} \delta R-\frac{r \bar{A_{0}}'}{\bar{N}\bar{h}} \delta A',
\end{eqnarray}
\begin{eqnarray}\label{41}
\frac{(1+\bar{f}_{R})\delta \dot{N}}{\bar{N}}=-\left(\frac{\bar{\psi}}{2 r}\right)'(\delta \dot{\psi}+\delta \dot{\psi^\ast})+\frac{i q \bar{A_{0}} \bar{\psi}'}{2r}(\delta \psi-\delta \psi^\ast)-\frac{i q \bar{A_{0}} \bar{\psi}}{2r}(\delta \psi'-\delta \psi'^\ast),
\end{eqnarray}
\begin{eqnarray}\label{42}
&&\bar{N} \delta A''+\bar{N}\left(\frac{2}{r}-\frac{\bar{h}'}{2\bar{h}}\right) \delta A'-\frac{q^2 \bar{\psi}^2}{r^2}\delta A=-\frac{q^2 \bar{A_{0}} \bar{\psi}^2}{r^2 \bar{\xi}}\delta \xi+\frac{\bar{N}\bar{A_{0}}'}{2\bar{h}}\delta h'+\frac{1}{2\bar{h}}\left[\frac{q^2\bar{A_{0}}\bar{\psi}^2}{r^2}-\frac{\bar{N}\bar{A_{0}'}\bar{h'}}{\bar{h}}\right]\delta h+\nonumber \\
&&\frac{i q  \bar{\psi}}{2r^2}(\delta \dot{\psi}-\delta \dot{\psi^\ast})+\frac{q^2 \bar{A_{0}} \bar{\psi}}{r^2}(\delta \psi+\delta \psi^\ast)
\end{eqnarray}
\begin{eqnarray}\label{43}
 \frac{\delta \dot{ A'}}{\sqrt{\bar{h}}}-\frac{\bar{A_{0}}'\delta \dot{ h}}{2 \bar{h}\sqrt{\bar{h}}}=\frac{i q  \bar{\psi} \bar{\xi}}{2 r^2}(\delta{\psi'}-\delta{\psi'^\ast})-\frac{i q \bar{\xi} \bar{\psi}'}{2r^2}(\delta \psi-\delta \psi^\ast),
\end{eqnarray}
\begin{eqnarray}\label{44}
&&-\delta \ddot{\psi}+\bar{\xi}^2 \delta \psi''+2iq \bar{A_{0}} \delta \dot{\psi}+ \bar{\xi} \bar{\xi}' \delta \psi'+\left(q^2 \bar{A_{0}}^2-\frac{\bar{\xi}\bar{\xi}'}{r}\right)\delta \psi-\frac{iq \bar{A_{0}} \bar{\psi}}{\bar{\xi}} \delta \dot{\xi}+\left(\bar{\xi} \bar{\psi}'-\frac{\bar{\xi} \bar{\psi}}{r}\right)\delta \xi'+iq \bar{\psi} \delta \dot{A}\nonumber\\
&&+\left(2 \bar{\xi} \bar{\psi}''+\bar{\xi}'\bar{\psi}'-\frac{\bar{\psi} \bar{\xi}'}{r}\right)\delta \xi+2 q^2 \bar{\psi}\bar{A_{0}} \delta A=0.
\end{eqnarray}
The perturbed scalar field is a complex quantity, written as $\delta \psi=\delta u+i \delta \dot{w}$. For the real part, we have
\begin{eqnarray}\label{45}
&&-\delta \ddot{u}+\bar{\xi}^2\delta u''-2q \bar{A_{0}}\delta \ddot{w}+\bar{\xi}\bar{\xi}'\delta u'+\left(q^2\bar{A_{0}}^2-\frac{\bar{\xi}\bar{\xi}'}{r}\right)\delta u+\left(\bar{\xi}\bar{\psi}'-\frac{\bar{\xi}\bar{\psi}}{r}\right)\delta \xi'+ \left(2\bar{\xi} \bar{\psi}''+\bar{\xi}'\bar{\psi}'-\frac{\bar{\psi} \bar{\xi}'}{r}\right)\nonumber\\
&&\times \delta \xi+2q^2\bar{\psi}\bar{A_{0}} \delta A=0,
\end{eqnarray}
and the imaginary part takes on the form
\begin{eqnarray}\label{46}
-\delta \dddot{w}+\bar{\xi}^2 \delta \dot{w}''+2q \bar{A_{0}} \delta \dot{u}+ \bar{\xi} \bar{\xi}' \delta \dot{w}'+\left(q^2 \bar{A_{0}}^2-\frac{\bar{\xi}\bar{\xi}'}{r}\right)\delta \dot{w}-\frac{q \bar{A_{0}} \bar{\psi}}{\bar{\xi}} \delta \dot{\xi}+q \bar{\psi} \delta \dot{A}=0.
\end{eqnarray}
Integration with respect to time of equations  (\ref{41}, \ref{43}) yields
\begin{eqnarray}\label{47}
(1+\bar{f_{R}})\frac{\delta N}{\bar{N}}=-\left(\frac{\bar{\psi}}{r}\right)'\delta u-\frac{ q \bar{A_{0}} \bar{\psi}'}{r}\delta w+\frac{q \bar{A_{0}} \bar{\psi}}{r}\delta w'+\delta \emph{F(r)},
\end{eqnarray}
\begin{eqnarray}\label{48}
\frac{\delta h}{\bar{h} \sqrt{\bar{h}}}= \frac{2\delta  A'}{\bar{A_{0}}'\sqrt{\bar{h}}}+\frac{ 2 q  \bar{\psi} \bar{\xi}}{r^2 \bar{A_{0}}'}\delta w'-\frac{ 2 q \bar{\xi} \bar{\psi}'}{\bar{A_{0}}' r^2}\delta w+\delta \textit{g(r)},
\end{eqnarray}
where $\delta \emph{F(r)}$ and $\delta \emph{g(r)}$ are arbitrary functions of the radial coordinate. Integration of equation (\ref{46}) now yields
\begin{eqnarray}\label{49}
\delta \ddot{w}-\bar{\xi}^2 \delta w''-2q \bar{A_{0}} \delta u-\bar{\xi} \bar{\xi}' \delta w'-\left(q^2 \bar{A_{0}}^2-\frac{\bar{\xi}\bar{\xi}'}{r}\right)\delta w+\frac{q \bar{A_{0}} \bar{\psi}}{\bar{\xi}} \delta \xi-q \bar{\psi} \delta A+\delta \textit{H(r)}=0,
\end{eqnarray}
where $\delta \textit{H(r)}$ is an arbitrary function of the radial coordinate. If an arbitrary function of $r$ is added to $\delta w$, due to the form of $\delta \psi$, it does not change which gives us the freedom to set $\delta \textit{H(r)}=0$.

A first-order differential equation is constructed from (\ref{38}) and (\ref{42}), using equations (\ref{47} -\ref{49})
\begin{eqnarray}\label{49a}
&&\delta \emph{F(r)}'+\left(\frac{\bar{\xi}'}{\bar{\xi}}+\frac{1}{r}\right)\delta \emph{F(r)}+\bar{f}_{RR}\left(\frac{1}{r}-\frac{1}{\bar{N}r}+\frac{r\bar{R}}{2 \bar{N}}+\frac{\bar{N}'}{\bar{N}}\right) \delta R=\frac{r \bar{A_{0}}\bar{A_{0}}'\bar{N}}{\bar{\xi}^2}(\sqrt{\bar{h}} \delta \emph{g(r)})'+\nonumber\\
&&\frac{r \bar{A'_{0}}^2}{2\bar{\xi}}\delta \emph{g(r)}+\frac{q^2\bar{A_{0}}^2\bar{\psi}^2\sqrt{\bar{h}}}{2r\bar{\xi}^2}\delta \emph{g(r)}.
\end{eqnarray}
Another first-order differential equation is obtained by plugging (\ref{38}) and (\ref{42}) in (\ref{47} )
\begin{equation}\label{49b}
\delta \emph{F(r)}'+\left(\frac{\bar{\xi}'}{\bar{\xi}}+\frac{1}{r}\right)\delta \emph{F(r)}+\bar{f}_{RR}\left(\frac{1}{r}-\frac{1}{\bar{N}r}+\frac{r\bar{R}}{2 \bar{N}}+\frac{\bar{N}'}{\bar{N}}\right) \delta R=\frac{r \bar{A'_{0}}^2}{2\bar{\xi}}\delta \emph{g(r)}.
\end{equation}
Substitution of (\ref{47}) and (\ref{48}) in (\ref{49}) then gives
\begin{eqnarray}\label{33}
&&\delta \ddot{w}-\bar{\xi}^2\delta w''+\left[\frac{q^2 \bar{A_{0}} \bar{\psi}^2}{r^2  \bar{A_{0}}'}\left(\frac{r \bar{A_{0}}' \bar{A_{0}}}{1+\bar{f}_{R}}+ \bar{N}\bar{h}\right)-\bar{\xi} \bar{\xi}'\right] \delta w'-\left[\frac{q^2 \bar{A_{0}} \bar{\psi} \bar{\psi}'}{r^2  \bar{A_{0}}'} \left(\frac{ r \bar{A_{0}} \bar{A_{0}}'}{1+\bar{f}_{R}}+\bar{N}\bar{h}\right)+q^2 \bar{A_{0}}^2\right.\nonumber \\
&&\left.-\frac{\bar{\xi} \bar{\xi}'}{r}\right] \times \delta w -q \bar{A_{0}}\left(2+\frac{\bar{\psi} \left(\frac{\bar{\psi}}{r}\right)'}{1+\bar{f}_{R}}\right)\delta u-q \bar{\psi} \delta A+\frac{q \bar{A_{0}} \bar{\psi}}{\bar{A_{0}}'} \delta A'+\frac{q \bar{A_{0}}\bar{\psi}\delta \emph{F(r)}}{1+\bar{f_{R}}}+\frac{q \bar{A_{0}}\sqrt{\bar{h}}\bar{\psi}\delta \emph{g(r)}}{2}\nonumber \\
&&=0,
\end{eqnarray}
Let us now substitute equations (\ref{40}, \ref{47} -\ref{49}, \ref{49b}) in (\ref{45}) and obtain
\begin{eqnarray}\label{50}
&&\delta \ddot{u}-\bar{\xi}^2 \delta u''-\bar{\xi} \bar{\xi}'\delta u'+\left[3 q^2 \bar{A_{0}}^2+\frac{\bar{\xi} \bar{\xi}'}{r}+\frac{ \bar{N}\left(\frac{\bar{\psi}}{r}\right)'^2 }{1+\bar{f}_{R}}\left(\frac{r^2 \bar{A_{0}}'^2}{2 (1+\bar{f}_{R})}-\frac{r^2 \bar{h} \bar{f}}{2( 1+\bar{f}_{R})}+\frac{r^2 \bar{h} \bar{f}_{R} \bar{R_{0}}}{2( 1+\bar{f}_{R})}-\bar{h}\right)\right]\nonumber \\
&& \times \delta u+2 q \bar{A_{0}} \bar{\xi}^2 \delta w''+q \bar{N}\bar{A_{0}}\left[2 \sqrt{\bar{h}}\bar{\xi}'+\frac{\bar{\psi} \left(\frac{\bar{\psi}}{r}\right)'}{1+\bar{f}_{R}}\left(\frac{\bar{h}}{r}-\frac{\bar{N}\bar{h} \bar{A_{0}}' }{\bar{A_{0}}}-\frac{r \bar{A_{0}}'^2}{2(1+\bar{f}_{R})}-\frac{\bar{h}\bar{f}_{R} \bar{R_{0}} r}{2 (1+\bar{f}_{R})}+\right.\right. \nonumber \\
&&\left.\left.\frac{\bar{h}\bar{f} r}{2 (1+\bar{f}_{R})}\right)\right]\delta w' +q \bar{A_{0}}\left[ 2 q^2 \bar{A_{0}}^2-\frac{2 \bar{\xi} \bar{\xi}'}{r}+\frac{\bar{\xi} \bar{\psi}' \left(\frac{\bar{\psi}}{r}\right)'}{1+\bar{f}_{R}}\left(\frac{\bar{\xi} \bar{A_{0}}'}{\bar{A_{0}}}-\bar{\xi}'-\frac{\bar{\xi}}{r }\right)\right]\delta w- \bar{\xi}^2 r \left(\frac{\bar{\psi}}{r}\right)'\times \nonumber\\
&&\frac{\delta \textit{F(r)}'}{1+\bar{f}_{R}}+\frac{\bar{f}_{RR}\bar{h}'}{2\left(1+\bar{f}_{R}\right)\bar{h}}\delta R=0.
\end{eqnarray}
Using equations (\ref{39}, \ref{48}, \ref{49}) then results in
\begin{eqnarray}\label{51}
&&\frac{q \bar{\psi}}{r}\left(\frac{\bar{A_{0}}}{1+\bar{f}_{R}}+\frac{\bar{N}\bar{h}}{r \bar{A_{0}}'}\right)\delta w''+ q \bar{A_{0}} \bar{\psi} \left[-\frac{q^2 \bar{h} \bar{\psi}^2}{r^4 \bar{A_{0}}'^2}+\frac{\bar{\xi}'}{(1+\bar{f}_{R}) \bar{\xi} r}+\frac{\sqrt{\bar{h}}\bar{\xi'}}{r^2 \bar{A_{0}} \bar{A_{0}}'}\right]\delta w'+\frac{q  \bar{A_{0}} \bar{\psi}}{r^2}\left[\frac{r q^2 \bar{A_{0}}^2}{(1+\bar{f}_{R}) \bar{\xi}^2}\right.\nonumber\\
&&\left.+\frac{q^2 \bar{A_{0}}}{\bar{N} \bar{A_{0}}'}-\frac{\bar{\xi}'}{(1+\bar{f}_{R}) \bar{\xi}}-\frac{\sqrt{\bar{h}}\bar{\xi}'}{r \bar{A_{0}} \bar{A_{0}}' }+\frac{q^2 \bar{h} \bar{\psi}\bar{\psi}'}{r^2 \bar{A_{0}}'^2}\right]\delta w-\frac{\left(\frac{\bar{\psi}}{r}\right)'}{1+\bar{f}_{R}}\delta u'-\left[ \frac{\left(\frac{\bar{\psi}}{r}\right)'}{1+\bar{f}_{R}}\left(\frac{1}{r}+\frac{\bar{\xi}'}{\bar{\xi}}\right) + \frac{\left(\frac{\bar{\psi}}{r}\right)''}{1+\bar{f}_{R}}\right]\delta u\nonumber\\
&&+\frac{\delta A''}{\bar{A_{0}}'} -\frac{\bar{A_{0}}''}{\bar{A_{0}}'^2}\delta A'+\frac{\bar{f}_{RR}\bar{h}'}{2\left(1+\bar{f}_{R}\right)\bar{h}}\delta R+\frac{\left(\sqrt{\bar{h}}\delta\textit{g(r)}\right)'}{2}=0.
\end{eqnarray}
Since ingoing boundary conditions (\ref{35}) must exist for all perturbations, including  perturbation of metric variables ($\delta N$ and $\delta h$) \cite{win}, equations (\ref{47}) and (\ref{48}) are needed to satisfy such conditions, leading to $\delta \textit{F(r)}=0$ and $\delta \textit{g(r)}=0$ at $r=r_{h}$. Therefore, $\delta \textit{F(r)}$, $\delta \textit{g(r)}$, $\delta \textit{F(r)}'$ and $\delta \textit{g(r)}'$ are removed from perturbation equations and $\delta R$ at $r=r_{h}$ must be zero on account of equations (\ref{49a}, \ref{49b}).

\end{document}